\documentclass[aps,prd,altaffilletter,twocolumn,superscriptaddress,floatfix,nofootinbib]{revtex4-2} 
\usepackage{amsfonts,amsmath,graphicx,bm}
\usepackage{tikz-feynman}
\usepackage{ifpdf}
\usepackage{multirow}
\usepackage{url}
\usepackage{todonotes}
\usepackage{float}
\usepackage{graphicx}
\usepackage{tablefootnote}
\usepackage{mathtools}
\usepackage[yyyymmdd,hhmmss]{datetime}
\usepackage{color}
\usepackage{bbold}
\usepackage{footmisc}
\usepackage{braket}
\usepackage{slashed}
\usepackage[colorlinks=true, linkcolor=red, urlcolor=red, citecolor=red]{hyperref}

\def\today{\number\day\space\ifcase\month\or
January\or February\or March\or April\or May\or June\or
July\or August\or September\or October\or November\or December\fi
\space\number\year}

\newcommand{\GeV}{\;\mathrm{GeV}}
\newcommand{\MeV}{\;\mathrm{MeV}}

\newcommand{\fm}{\;\mathrm{fm}}

\begin{document}
\title{Constraints on axion-like particles using lattice QCD calculations of the rate for  $J/\psi \to \gamma a$}

\author{Brian~\surname{Colquhoun}} 
\email[]{Brian.Colquhoun@glasgow.ac.uk}
\affiliation{School of Physics and Astronomy, University of Glasgow, Glasgow, G12 8QQ, UK}

\author{Christine~T.~H.~\surname{Davies}} 
\email[]{Christine.Davies@glasgow.ac.uk}
\affiliation{School of Physics and Astronomy, University of Glasgow, Glasgow, G12 8QQ, UK}

\author{G.~Peter~\surname{Lepage}} 
\affiliation{Laboratory of Elementary Particle Physics, Cornell University, Ithaca, New York 14853, USA}

\author{Sophie~\surname{Renner}} 
\email[]{Sophie.Renner@glasgow.ac.uk}
\affiliation{School of Physics and Astronomy, University of Glasgow, Glasgow, G12 8QQ, UK}

\collaboration{HPQCD Collaboration}
\homepage[URL: ]{https://www.physics.gla.ac.uk/hpqcd/}

\begin{abstract}
\noindent A key search mode for axion-like particles (ALPs) that couple to charm quarks is $J/\psi \to \gamma a$. Here we calculate the form factor that allows the rate of this process to be determined using lattice QCD for the first time. Our calculations use the relativistic Highly Improved Staggered Quark (HISQ) action for the valence charm quarks on gluon field configurations generated by the MILC collaboration that include $u$, $d$, $s$ and $c$ HISQ quarks in the sea at four values of the lattice spacing and both unphysical and physical sea quark masses. We determine the form factor as a function of ALP mass with an uncertainty of less than 2\% across our full range of ALP masses from zero up to 95\% of the $J/\psi$ mass. This represents a substantial improvement in accuracy of the theoretical picture of this decay compared to the previously used tree-level and $\mathcal{O}(\alpha_s)$ perturbation theory. We use our form factor to determine constraints on ALP masses and couplings to charm quarks and photons in several different scenarios using recent experimental data from BESIII. Our calculation paves the way for further lattice QCD input on new physics constraints from radiative decays. 

\end{abstract}

\maketitle


\section{Introduction}
\label{sec:intro}

Axions and axion-like particles (ALPs) frequently appear in extensions of the Standard Model (SM) as pseudo Nambu-Goldstone bosons associated with spontaneous breaking of global symmetries.  Although the parameters of the QCD axion~\cite{Peccei:1977hh,Peccei:1977ur,Weinberg:1977ma,Wilczek:1977pj} introduced to solve the strong CP problem are highly constrained, light pseudoscalar relics of new physics can occupy a much broader parameter space. In general the ALP will couple to all SM particles with couplings suppressed, at leading-order, by one power of the scale of the global symmetry breaking, assumed to be much larger than the electroweak scale. The couplings will then be weak, and the impact of ALPs consequently small, for a heavy new physics sector. This means that several different, complementary, search modes are required that are each capable of relatively high precision and allow constraints to be placed under a variety of ALP scenarios. Astrophysical observations and beam dump experiments can probe very light (sub-MeV) ALP masses (see e.g.~\cite{ParticleDataGroup:2024cfk,AxionLimits}) and high-energy collider experiments can search for heavy (multi-GeV mass) ALPs (e.g.~\cite{Bauer:2017ris}), but for masses in the few MeV to few GeV range flavour physics observables provide the best constraints even if the underlying theory is flavour conserving (e.g.~\cite{Izaguirre:2016dfi,Gavela:2019wzg,Bauer:2021mvw}). It is this range that we will focus on here. 

It is important that the ALP constraints derived are reliable, using an accurate theoretical understanding of the process in the particular scenario being considered. 
For flavour physics processes involving hadron decays or hadron to hadron transitions, nonperturbative QCD effects are important. Calculations using lattice QCD are then needed to take such effects fully into account. 

For ALPs with masses in the range of $100\, \mathrm{MeV}/c^2$ to a few $\mathrm{GeV}/c^2$ an important search process is the radiative decay of a heavyonium vector meson $V$ produced, for example, in $e^+e^-$ annihilation. The decay $V\to \gamma a$, where $\gamma$ is a real photon and $a$ a real ALP, can proceed via two routes. In one the vector meson annihilates to a virtual photon that decays to $\gamma a$ via the $a\gamma\gamma$ coupling. In the second a photon is radiated from the valence quark or antiquark in $V$ and the resulting pseudoscalar heavyonium configuration annihilates to $a$ via the $aq\overline{q}$ coupling. The relative strength of the two amplitudes depends on the quark and photon couplings of the ALP but also on the hadronic quantities that parameterise the strong interaction physics inside $V$. In the first case this is the vector meson decay constant; in the second case this is a form factor that is a function of the photon virtuality and the ALP mass. 

Here we study the process $J/\psi \to \gamma a$ and provide the first results using lattice QCD for the form factor for this process via the $ac\overline{c}$ coupling. Previous ALP constraints from this process either set the $ac\overline{c}$ coupling to zero (for example~\cite{BESIII:2022rzz}), or used tree-level~\cite{Merlo:2019anv,DiLuzio:2024jip} or $\mathcal{O}(\alpha_s)$ NRQCD perturbation theory~\cite{Nason:1986tr,Carmona:2021seb,Bauer:2021mvw} to approximate the form factor with an expression relating it to the $J/\psi$ decay constant and mass. Our results provide instead the first fully nonperturbative analysis of the form factor, which allows us to quantify how well the perturbative approximation works. We are also able to provide the first accurate complete calculation of the rate for $J/\psi \to \gamma a$ as a function of ALP mass, combining the amplitudes that depend on the $a\gamma\gamma$ and $ac\overline{c}$ couplings. We use this along with recent experimental results from the BESIII collaboration to provide improved constraints on ALP parameters. 

The layout of the paper is as follows:  Section~\ref{sec:pert} defines the form factor that we will calculate in lattice QCD and derives the perturbative approximations that have been used for it previously; Section~\ref{sec:lattice} describes the lattice QCD calculation with subsection~\ref{sec:results} giving the results for the form factor in the continuum limit; Section~\ref{sec:constraints} derives constraints on ALP parameters combining the lattice QCD results for the expected rate with experimental search limits and Section~\ref{sec:conclusions} gives our conclusions and prospects for future calculations. 

\section{Comparing lattice QCD to the current perturbative approach}\label{sec:pert}

\begin{figure}
		\includegraphics[width=0.45\textwidth]{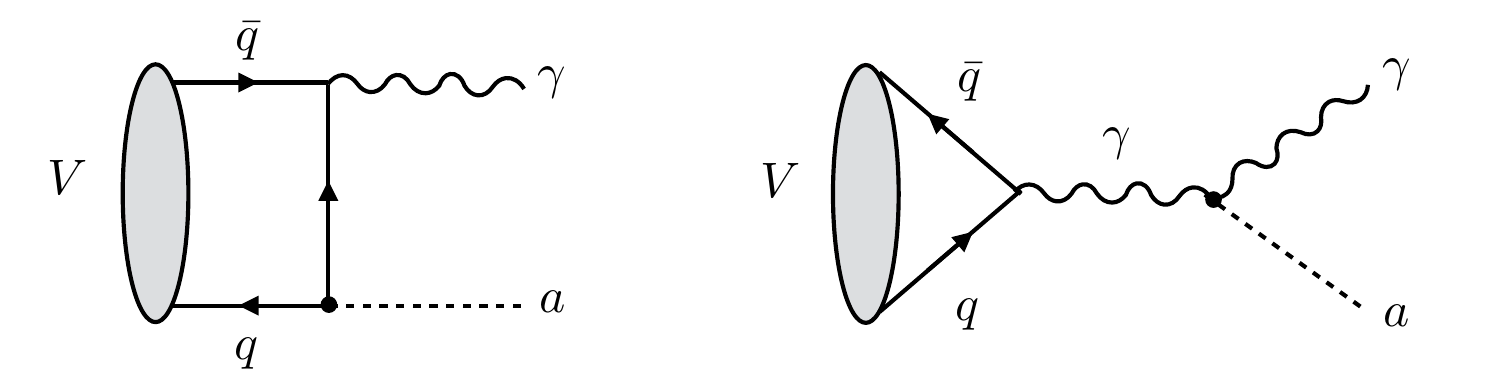}
		\caption{Diagrams contributing to the decay $V\to\gamma a$ for heavyonium vector meson $V$ with valence quarks $q\overline{q}$. The form factor parameterising the amplitude for the left-hand diagram in which the ALP couples to $q\overline{q}$ is calculated here for the process $J/\psi \to \gamma a$ using lattice QCD.}
		\label{fig:feyn-diags}
\end{figure}
The effective Lagrangian describing the interaction of an ALP with fermions and photons is 
\begin{equation}
\label{eq:L-int}
\mathcal{L} =\frac{\partial ^{\mu} a}{2f_a}\sum_f c_{ff}(\mu)\overline{f}\gamma_{\mu}\gamma_5 f + c_{\gamma\gamma}\frac{\alpha}{4\pi}\frac{a}{f_a}F_{\mu\nu}\tilde{F}^{\mu\nu} \,.
\end{equation}
Here the ALP energy scale $f_a$ (known as the ALP decay constant) is assumed to be independent of (and typically much larger than) its mass $m_a$ and $c_{ff}$ and $c_{\gamma\gamma}$ are dimensionless effective couplings. The scale dependence of these couplings will be discussed in Sec.~\ref{sec:constraints}. We can rewrite the first term of Eq.~\eqref{eq:L-int} using the anomaly equation for the divergence of the axial vector current to give~\cite{Bauer:2017ris}
\begin{equation}
\label{eq:L-int-rewrite}
-\sum_f\frac{m_fc_{ff}(\mu)}{f_a} a \overline{f} i \gamma_5 f + c_{ff}\frac{N_c^fQ_f^2e^2}{16\pi^2}\frac{a}{f_a}F_{\mu\nu}\tilde{F}^{\mu\nu} + \ldots \, ,
\end{equation}
The first term shows the ALP coupling to a pseudoscalar fermion current multiplied by the fermion mass, the second term gives a loop correction\footnote{This correction applies in the limit that $m_f \ll m_a$. In general there is a mass-dependent function in the coefficient which has the effect of decoupling heavier fermions, see Sec.~\ref{sec:constraints}.} to $c_{\gamma\gamma}$ and $\ldots$ gives similar terms for gluons and weak gauge bosons.

The two diagrams of Fig.~\ref{fig:feyn-diags} show the two routes for the process $V\to \gamma a$: on the left  via the $c_{ff}$ coupling (where $f$ is the flavour of the valence quark $q$ in $V$) and on the right via $c_{\gamma\gamma}$. Our lattice QCD calculation provides an improved analysis for the diagram on the left for $J/\psi \to \gamma a$ with $q$ a charm quark, and we first discuss that case. 

As mentioned in Section~\ref{sec:intro}, the processes of Fig.~\ref{fig:feyn-diags} depend on the internal structure of the $J/\psi$ meson, which is determined by nonperturbative strong-interaction physics. Appropriate hadronic parameters, calculable in lattice QCD, should be defined for each diagram. We define the form factor $F$ for the left-hand diagram of Fig.~\ref{fig:feyn-diags} from the matrix element, $\mathcal{M}$, by\footnote{Symmetry considerations show there is only a single form factor, as for the process $J/\psi\to \gamma \eta_c$~\cite{Dudek:2006ej}.}  
\begin{eqnarray}
\label{eq:Fdef}
\mathcal{M}(J/\psi\to \gamma a) &=& \frac{2F(q^2,m_a)}{M_{J/\psi}}\varepsilon^{\alpha\beta\delta\epsilon}p_{\alpha}p^{\prime}_{\beta}\varepsilon^{J/\psi}_{\delta}\varepsilon^{\gamma}_{\epsilon} \times \nonumber \\
&& \hspace{4.0em} \left( \frac{m_c c_{cc}}{f_a}eQ_c\right)
\end{eqnarray}
where $p$ is the 4-momentum of the $J/\psi$, $p^{\prime}$ that of the ALP and  $q=p-p^{\prime}$. The dimensionless form factor $F$ is a function of photon virtuality, $q^2$, and the ALP mass. Here we will only consider real photons with $q^2=0$. $\varepsilon^{J/\psi}$ and $\varepsilon^{\gamma}$ are the polarisation vectors of the $J/\psi$ and photon respectively. The factor of 2 in Eq.~\eqref{eq:Fdef} allows for the fact that there are two diagrams each giving the same contribution. Only one diagram  is drawn on the left in Fig.~\ref{fig:feyn-diags} and we will only calculate one correlation function on the lattice. The second diagram has the photon and ALP interchanged. 

The width for the decay to a real photon-ALP pair is then 
\begin{eqnarray}
\label{eq:width-cc}
\Gamma &=& \frac{c_{cc}^2}{f_a^2}\frac{\alpha Q_c^2} {6} m_c^2 [F(0,m_a)]^2 M_{J/\psi}\left(1-\frac{m_a^2}{M_{J/\psi}^2}\right)^3 \nonumber \\
&=& \frac{\alpha Q_c^2M_{J/\psi}^3}{24f_a^2} \left(1-\frac{m_a^2}{M_{J/\psi}^2}\right) \left| c_{cc}\tilde{F} \right|^2                                                     \, ,
\end{eqnarray}
where 
\begin{equation}
\label{eq:Ftildedef}
\tilde{F} \equiv \frac{2m_c}{M_{J/\psi}} F(0,m_a)\left(1-\frac{m_a^2}{M^2_{J/\psi}}\right)  \, 
\end{equation}
is introduced to facilitate the comparison with perturbation theory.
 
 $\tilde{F}$ is known from NRQCD perturbation theory through lowest and first order~\cite{Wilczek:1977zn,Nason:1986tr,Bauer:2021mvw}:
 \begin{equation}
\label{eq:Ftildepert}
\tilde{F}_{\mathrm{pert,NLO}} = 2\left|\psi(0)\right|\sqrt{\frac{12}{M_{J/\psi}^3}}\left(1-\frac{2\alpha_s(\mu)}{3\pi}a_P(x)+\mathcal{O}(\alpha_s^2) \right) 
\end{equation}
 where $|\psi(0)|$ is the $J/\psi$ nonrelativistic wavefunction at the origin and $x=1-m_a^2/M_{J/\psi}^2$. $a_P$ varies from $a_P(0)=2$ to $a_P(1)\approx 6.62$~\cite{Bauer:2021mvw} and $\alpha_s \approx 0.25$ at scale $\mu=M_{J/\psi}$, giving substantial first order corrections at small $m_a$. Note that $m_c$ does not appear in this expression because it has been set equal to $M_{J/\psi}/2$ in leading-order NRQCD. 
 
 The wavefunction at the origin can be estimated from the decay rate for $J/\psi \to e^+e^-$ which has also been calculated through first order in perturbation theory~\cite{Barbieri:1975ki}:\footnote{Note that there is a typographical error of a factor of 16  in Refs~\cite{Nason:1986tr,Bauer:2021mvw} in reporting this result.} 
 \begin{eqnarray}
\label{eq:Vtoee}
\Gamma(J/\psi \to e^+e^-) &\equiv& \frac{4\pi}{3}\alpha^2Q_c^2\frac{f_{J/\psi}^2}{M_{J/\psi}}  \\
\label{eq:Vtoee-nonrel}
&& \hspace{-6.0em}= 16\pi\alpha^2Q_c^2\frac{\left| \psi(0) \right|^2}{M^2_{J/\psi}} \left( 1-\frac{16\alpha_s}{3\pi} + \mathcal{O}(\alpha_s^2)\right) \, . 
\end{eqnarray}
 This equation allows us to express the wavefunction at the origin in terms of the $J/\psi$ decay constant, $f_{J/\psi}$:
 \begin{equation}
 \label{eq:psi0}
 \left|\psi(0)\right| = f_{J/\psi}\sqrt{\frac{M_{J/\psi}}{12}}\left(1+\frac{8\alpha_s}{3\pi} + \ldots \right) \, .
 \end{equation}
 The decay constant can be determined from the measured leptonic decay rate of the $J/\psi$ or, more accurately, from lattice QCD calculations~\cite{Hatton:2020qhk}. Substituting Eq.~\eqref{eq:psi0} into Eq.~\eqref{eq:Ftildepert} for $\tilde{F}_{\mathrm{pert,NLO}}$ gives us finally
  \begin{equation}
\label{eq:Ftildepert2}
\tilde{F}_{\mathrm{pert,NLO}} = 2\frac{f_{J/\psi}}{M_{J/\psi}}\left(1-\frac{2\alpha_s(\mu)}{3\pi}b_P(x)+ \ldots\right)  \, ,
\end{equation}
where $b_P\equiv a_P(x)-4$ varies between $b_P(0) = -2$ and $b_P(1) = 2.62$. The first order correction given by $b_P$ is then smaller than that given by $a_P$ (at small $m_a$) but, as we have seen, it results from the combination of two large corrections so the implications for further corrections to the perturbative result at higher order in $\alpha_s$ are not clear. 

The leading order result above for $\tilde{F}_{\mathrm{pert,NLO}}$ is $2f_{J/\psi}/M_{J/\psi}=0.265(1)$ using the lattice QCD result for $f_{J/\psi}$ from~\cite{Hatton:2020qhk} and the experimental $J/\psi$ mass~\cite{ParticleDataGroup:2024cfk}. Note that this result has no dependence on $m_a$. $\tilde{F}_{\mathrm{pert,NLO}}$ develops dependence on $m_a$ through $b_P(x)$.
 In Section~\ref{sec:lattice} we will examine, by comparison to the lattice QCD results of Eq.~\eqref{eq:Ftildedef}, how well the perturbative approach works, both without and with the $\mathcal{O}(\alpha_s)$ corrections. We might expect the perturbative approach to work best at small values of $m_a$, where momenta flowing inside the perturbative diagram are relatively high and well away from bound-state effects that will appear as $m_a$ approaches the mass of the $\eta_c$~\cite{Polchinski:1984ag,Pantaleone:1984ug}. 
 
 A further issue for the perturbative approach is that of relativistic corrections. In principle these could be of $\mathcal{O}(v^2/c^2)\approx 30\%$ for the $J/\psi$, where $v^2/c^2$ is the squared velocity of the quark inside the meson. The largest such corrections, however, are those that affect $\psi(0)$ and these will be removed when $\psi(0)$ is expressed in terms of $f_{J/\psi}$. The remaining corrections should be significantly smaller. Such corrections have been assessed for the similar process $\eta_c \to \gamma\gamma$ using lattice QCD calculations~\cite{Colquhoun:2024wsj} 
and again taking a ratio of the form factor to the decay constant (calculated in lattice QCD) to compare to expectations from leading-order nonrelativistic QCD. By studying both $\eta_c$ and $\eta_b$, which have very different values for $v^2/c^2$, it can be seen that the relativistic corrections are much smaller than might initially have been expected. 

We now return to the calculation of the full rate for $J/\psi \to \gamma a$. The righthand diagram of Fig.~\ref{fig:feyn-diags} is straightforward to analyse, because the annihilation of $V$, governed by QCD, factorises from the interactions of the virtual photon produced. The amplitude can therefore be simply expressed in terms of the $J/\psi$ decay constant, by definition of the decay constant (see Eq.~\eqref{eq:Vtoee} for the $J/\psi \to e^+e^-$ case). The total rate for $J/\psi \to \gamma a$ is then given by  
\begin{eqnarray}
\label{eq:full-rate}
\Gamma(J/\psi \to \gamma a) &=& \frac{\alpha Q_c^2 M_{J/\psi}^3}{24f_a^2}\left(1-\frac{m_a^2}{M_{J/\psi}^2}\right)\times \\
&&\left| c_{cc}(\mu_c)\tilde{F} - c_{\gamma\gamma}\frac{\alpha}{\pi}\frac{f_{J/\psi}}{M_{J/\psi}}\left(1-\frac{m_a^2}{M_{J/\psi}^2}\right)\right|^2 \nonumber \, .
\end{eqnarray} 
In Sec.~\ref{sec:constraints} we will use this expression, with the lattice QCD results for $\tilde{F}$ and $f_{J/\psi}$, to determine constraints on ALP parameters from recent experimental searches.

\section{The lattice calculation} \label{sec:lattice}

We determine the matrix element for $J/\psi \to \gamma a$ (Eq.~\eqref{eq:Fdef}) from the correlation function between an operator that creates a charmonium state at rest with vector quantum numbers and currents that couple to an ALP and a photon (with chosen values of equal and opposite spatial momentum). The correlation function is constructed from charm quark propagators calculated as solutions of the Dirac equation on background gluon field configurations generated with the appropriate probability distribution for the Feynman Path Integral. So far this is a standard lattice QCD calculation of a `3-point' correlation function. At the next stage we use a technique from the study of the process $\eta_c \to \gamma\gamma$~\cite{Dudek:2006ut,Colquhoun:2023zbc}. We imagine coupling an external photon to the correlation function and form a weighted sum over the time insertion point of the electromagnetic current to set the photon to a fixed energy, so that the photon is on-shell. This then sets the energy and momentum of the external ALP to those corresponding to a chosen mass. Fitting the resulting `2-point' correlation function, along with a standard vector charmonium 2-point correlator, allows us to extract the matrix element, and form factor, that we need here. By combining results as a function of ALP mass on gluon field configurations corresponding to a range of different values of the lattice spacing we can determine the form factor in the continuum (zero lattice spacing) physical limit. 

Those uninterested in the details of the lattice QCD calculation should skip to subsection~\ref{sec:results} where we show the final lattice QCD results for the $J/\psi \to \gamma a$ form factor. 

\subsection{Calculational Details} \label{sec:details}

We use the Highly Improved Staggered Quark (HISQ) formalism~\cite{Follana:2006rc} for all the quarks in this calculation. The HISQ discretisation of the Dirac action is improved to remove all discretisation errors at $\mathcal{O}(a^2)$, where $a$ is the lattice spacing. This leaves discretisation errors at higher order in $a$ (i.e. $\mathcal{O}(a^4)$) and those generated at $a^2$ by radiative corrections. The high level of improvement and consequent small size of discretisation errors is important for calculations, such as this one,  involving heavy charm quarks because discretisation errors can be inflated by the presence of the relatively large scale of the charm quark mass. The HISQ action includes an improved discretisation of the Dirac derivative using a 3-link `Naik' term. For the charm quarks (both sea and valence) we use a coefficient for the Naik term~\cite{Follana:2006rc,Monahan:2012dq} that further improves the action by removing tree-level discretisation errors at $\mathcal{O}((am_c)^4)$. 

We work on gluon field configurations that include $u$, $d$, $s$ and $c$ HISQ quarks in the sea generated by the MILC collaboration~\cite{Bazavov:2010ru,Bazavov:2012xda}, with parameters given in Table~\ref{tab:params}. The different sets of gluon field configurations have values for the lattice spacing ranging from 0.15 fm to 0.06 fm. This wide range of values allows us to assess the discretisation errors and remove them accurately when we take the $a\to 0$ continuum limit in Section~\ref{sec:continuum}. Table~\ref{tab:params} gives the masses of the sea quark masses included in the gluon field configurations. The $s$ and $c$ sea quark masses are close to their physical values on all sets but the $u$ and $d$  quark masses (taken to be the same) cover a range. Most ensembles have the $u/d$ sea mass unphysically heavy at one-fifth of the corresponding $s$ quark mass but two have a value close to the physical value for the $u/d$ average of $m_s/27$. Again this range allows us to assess the impact of the $u/d$ sea quark mass on the results so that we can also take all sea masses to their physical values along with the continuum extrapolation.  

Table~\ref{tab:params} also gives the valence $c$ quark masses used on each set. These were determined in~\cite{Hatton:2020qhk} so that the $J/\psi$ mass determined on each set of gluon field configurations is equal to its experimental value~\cite{ParticleDataGroup:2024cfk}. The values for the Naik coefficient for each valence $c$ mass are also given in~\cite{Hatton:2020qhk}. The sea $c$ mass is not tuned as accurately and so differs slightly from the valence $c$ mass. We expect the sea $c$ quarks to have very little effect on the form factor so this is not a significant issue. We will in any case include a term to correct for the mistuning of the sea $c$ mass when we fit the lattice results to obtain the form factor at the physical point in Section~\ref{sec:continuum}.

\begin{table}
\caption{Parameters of the gluon field configurations~\cite{Bazavov:2010ru,Bazavov:2012xda} that we use. The lattice spacing is given in the second column in units of $w_0$~\cite{Borsanyi:2012zs} where $w_0=0.1715(9)\fm$ is determined from lattice QCD calculations of the pion decay constant, $f_{\pi}$~\cite{Dowdall:2013rya}.
Sets  labelled `vc' are `very coarse' ($a \approx 0.15\fm$), `c' are `coarse' ($a \approx 0.12\fm$), `f' are `fine' ($a \approx 0.09\fm$), and `sf' are
`superfine' ($a \approx 0.06\fm$).  `5' or `phys' refer to the ratio $m_s^{\mathrm{sea}}/m_l^{\mathrm{sea}}$ (the physical ratio is $\approx 27$).
The HISQ action is used for both the sea quark masses, $m_{l,s,c}^{\mathrm{sea}}$ ($m_u=m_d=m_l$) and the valence charm quark masses, $am_c^{\mathrm{val}}$. Their masses are expressed in lattice units.
$L_s$ and $L_t$  give the extent of the lattice in spatial and time directions. Sets denoted vc-5, c-5, c-phys and f-5 include 1000 gluon field configurations each and we used two different positions for the time origin to increase statistical precision. On f-phys we used 200 gluon field configurations and on sf-5, 500, with a single time origin.  
}
\begin{tabular}{llllllllll}
\hline
\hline
Set & $w_0/a$ & $am_{l}^{\mathrm{sea}}$ & $am_{s}^{\mathrm{sea}}$ & $am_{c}^{\mathrm{sea}}$ & $am_c^{\mathrm{val}}$ & $L_s$ & $L_t$ \\
\hline
vc-5 & 1.1119(10) & 0.013 & 0.065 & 0.838 & 0.888 & 16 & 48 \\
\hline
 c-5 & 1.3826(11) & 0.0102 & 0.0509 & 0.635 & 0.664 & 24 & 64  \\
 c-phys & 1.4149(6) & 0.00184 & 0.0507 & 0.628 & 0.643 & 48 & 64 \\
 \hline
 f-5 & 1.9006(20) & 0.0074 & 0.037 & 0.440 & 0.450 & 32 & 96  \\
 f-phys & 1.9518(7) & 0.00120 & 0.0363 & 0.432 & 0.433 & 64 & 96 \\
\hline
 sf-5 & 2.8960(60) & 0.00480 & 0.0240 & 0.286 & 0.274 & 48  & 144 \\
\hline
\hline
\end{tabular}
\label{tab:params}
\end{table}

\begin{figure}
		\includegraphics[width=0.4\textwidth]{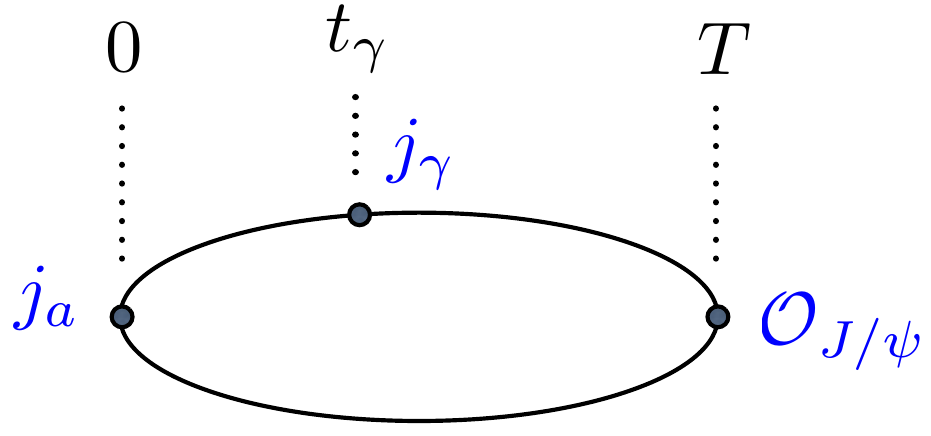}
		\caption{Schematic diagram of the connected 3-point correlation function between $O_{J/\psi}$ and a vector current, $j_{\gamma}$ and pseudoscalar current, $j_a$. The lines between the operators represent $c$ quark propagators. We do not include any quark-line disconnected correlation functions in our calculation. }
		\label{fig:3pt-pic}
\end{figure}
Propagators for the valence $c$ quarks are calculated on each gluon field configuration and then combined into correlation functions, inserting appropriate operators at different time points and summing over spatial positions. The simplest correlation function is the vector charmonium `2-point' correlator that creates charmonium states with vector quantum numbers at $t=0$ and destroys them at $t=t$: 
\begin{equation}
\label{eq:C2pt}
C_{2pt}(t) = \langle 0 | \mathcal{O}_{J/\psi}(0) \mathcal{O}^{\dag}_{J/\psi}(t) | 0\rangle \, . 
\end{equation}
$\mathcal{O}_{J/\psi}$ is a staggered-quark implementation of $\overline{c}\gamma_x c$ and we sum over spatial positions at $t=0$ (using a random wall source for the propagators) and at $t=t$ to set the vector charmonium states at rest on the lattice. All values of $t$ on the lattice are obtained in a single calculation. The $J/\psi$ is the lightest state that will appear in this correlation function. 

The 3-point correlation function that we need for $J/\psi \to \gamma a$,
\begin{equation}
\label{eq:C3pt}
C_{3pt}(t_{\gamma},T) = \langle 0 | j_a(0) j_{\gamma}(t_{\gamma})\mathcal{O}^{\dag}_{J/\psi}(T) | 0\rangle \, , 
\end{equation}
 is shown schematically in Fig.~\ref{fig:3pt-pic}. This now includes operators $\mathcal{O}_{J/\psi}$ to create vector charmonium states, $j_{\gamma}$ to couple to an external photon and $j_a$ to couple to an external ALP. $j_{\gamma}$ is a staggered-quark implementation of $\overline{c}\gamma_z c$ (without the electric charge factor of $eQ_c$) and $j_a$, of $\overline{c}\gamma_5 c$. In staggered quark spin-taste notation we use the point-split operator $\gamma_x\gamma_t \otimes \gamma_5\gamma_z$ for $\mathcal{O}_{J/\psi}$, the local $\gamma_z\otimes \gamma_z$ operator for $j_{\gamma}$ and the local $\gamma_5\otimes\gamma_5$ operator for $j_a$. This configuration of spin-tastes was previously used to calculate the form factor for $J/\psi \to \gamma \eta_c$~\cite{Colquhoun:2023zbc}. Note that $\mathcal{O}_{J/\psi}$ and $j_{\gamma}$ have orthogonal spin polarisations.

\begin{table}[!h]
\caption{Values of $T$ used on each ensemble. In each case there are additional values on the other side of the time origin, i.e. at $-T$. }
\label{tab:t_values}
\begin{tabular}{ll}
\hline
\hline
Set & $T$ \\
\hline
vc-5 & 10, 11, 12, 13, 14, 15\\
vc-5 ($m_a=2.5\GeV$) & 15, 16, 17, 18, 19, 20\\
\hline
 c-5 & 10, 13, 16, 19   \\
 c-phys & 10, 13, 16   \\
 \hline
 f-5    & 16, 19, 22, 25 \\
 f-phys & 16, 19, 22, 25 \\
\hline
 sf-5 & 24, 29, 34, 39  \\
\hline
\hline
\end{tabular}
\end{table}

It is convenient to construct the 3-point correlation function by positioning the pseudoscalar current at the time origin as in Fig.~\ref{fig:3pt-pic}. A $c$ quark propagator with source at the origin is then used as the source, at $T$, for another propagator using the standard sequential source technique. This second propagator is then tied up, at $t_{\gamma}$, with another $c$ quark propagator from the origin to make the 3-point correlator. Each value of $T$ requires a separate calculation. We use 6, 8 or 12 values of $T$, split evenly both sides of the time origin, and ranging from $\approx 1.2\fm$ upwards (see~\cite{Colquhoun:2023zbc} (Appendix B) for a test of this approach). The $T$ values are listed in Table~\ref{tab:t_values} (for one side of the origin only).

To achieve the physical configuration in which the photon and ALP have equal and opposite spatial momentum, we introduce momentum into the $c$ quark propagator connecting $t=0$ and $t=t_{\gamma}$. This is done using twisted boundary conditions~\cite{Sachrajda:2004mi,Guadagnoli:2005be}. The magnitude of the spatial momentum, $|\mathbf{q}|$, that we need is 
\begin{equation}
\label{eq:momval}
  |\mathbf{q}| = \frac{M_{J/\psi}}{2}\left(1-\frac{m^2_a}{M^2_{J/\psi}}\right)
\end{equation}
for an on-shell photon and ALP.  To cover the range of accessible ALP masses we choose $m_a$ values to be $0,1,2$ and $2.5\GeV$. We evaluate the $ |\mathbf{q}|$ values needed using the experimental value of the $J/\psi$ mass. The small difference between the $J/\psi$ mass from experiment and its value as measured on each lattice ensemble means the values of $m_a$ are slightly different from our four target values.

The spatial momentum is inserted using a twist (phase) applied to the gluon fields. Since we have chosen polarisations $x$ and $z$ for the $J/\psi$ and photon, Eq.~\eqref{eq:Fdef} shows that we need a component of the spatial momentum to be in the $y$ direction. In fact we simply take the spatial momentum to be entirely in the $y$-direction (see supplementary material of~\cite{Colquhoun:2024wsj} for a test of a different momentum arrangement). The twist $\theta$ is then set as
\begin{equation}
\label{eq:theta}
\mathbf{\theta} = (0,\theta,0);  \quad \theta=\frac{a|\mathbf{q}|L_s}{\pi} \,.
\end{equation}
Values of $\theta$ used on each ensemble are listed in Table~\ref{tab:twists}.

\begin{table}[!h]
  \caption{Twist, $\theta$, used for each momentum insertion for our target ALP masses of $0\GeV$, $1\GeV$, $2\GeV$ and $2.5\GeV$. On set sf-5, we have included two additional $\theta$ values of 6.3066 and 3.7685 from a slightly mistuned test calculation.}
\label{tab:twists}
\begin{tabular}{c|cccc}
\hline
\hline
\multirow{2}{*}{Set}&\multicolumn{4}{c}{Twist, $\theta$}\\
    \cline{2-5}
 & $m_a=0\GeV$ & $1\GeV$ & $2\GeV$ & $2.5\GeV$ \\
\hline
vc-5 & 6.1642 & 5.5215 & 3.5933 & 2.1472 \\
\hline
 c-5  &  7.4360 & 6.6607 & 4.3347 & 2.5902   \\
 c-phys & 14.5325 & 13.0172 & 8.4715 & 5.0622  \\
 \hline
 f-5    & 7.2125 & 6.4605 & 4.2044 & 2.5123 \\
 f-phys & 14.0465 & 12.5820 & 8.1882 & 4.8929 \\
\hline
 sf-5 & 7.1048 & 6.3640 & 4.1416 & 2.4748 \\
\hline
\hline
\end{tabular}
\end{table}

The 3-point correlation function $C_{3pt}(t_{\gamma},T)$ is then converted to a 2-point function, $\tilde{C}_{2pt}$, between $0$ and $T$ by performing a weighted sum over all values of $t_{\gamma}$ from $-L_t/2+1$ to $+L_t/2-1$. This sum, with its weighting of $\exp(-\omega t_{\gamma})$, sets the photon energy to $\omega$~\cite{Ji:2001wha,Ji:2001nf,Dudek:2006ej,Colquhoun:2023zbc}. For our kinematics, $\omega=|\mathbf{q}|$. We have
\begin{equation}
\label{eq:3to2}
\tilde{C}_{2pt}(T)= a \sum_{t_{\gamma}} e^{-\omega t_{\gamma}} C_{3pt}(t_{\gamma},T) \, .
\end{equation}
The summand is strongly peaked around $t_{\gamma}=0$ (i.e. where $j_{\gamma}$ and $j_a$ are close together in time) and $T$ values are chosen to lie outside this peak~\cite{Colquhoun:2023zbc}. This results in a good projection onto the ground-state $J/\psi$ meson and a large time separation between the $J/\psi$ and the pseudoscalar and vector currents. This enables us to extract the matrix element for $J/\psi \to \gamma a$ cleanly from $\tilde{C}_{2pt}(T)$. For further discussion on this point, see Appendix~\ref{appendix:reach}. This shows that, for ALP masses up to about $0.9 M_{J/\psi}$ the summand for Eq.~\eqref{eq:3to2} is well-behaved as described above. For ALP masses very close to $M_{J/\psi}$ with very small photon energy, this picture breaks down, however. This provides a practical upper bound on the ALP mass we can use in the calculation of the form factor but this is close to the point where ALP/$\eta_c$ mixing, not allowed for in our calculation (or in the perturbation theory), would start to be an issue in any case.

\begin{table*}
  \caption{Raw lattice results for $aM_{J/\psi}$ and $\tilde{F}$ from our correlator fits for each set of gluon field configurations and each target value for ALP mass, $m_a$. On sf-5, the two additional points (at $m_a=1.04\,\mathrm{GeV}$ and 2.12 GeV) have $\tilde{F}=0.20706(66)$ and $0.22906(75)$.}
  \label{tab:rawresults}
  \begin{tabular}{c|c|cccc}
    \hline
    \hline
    \multirow{2}{*}{Set} & \multirow{2}{*}{$aM_{J/\psi}$} & \multicolumn{4}{c}{$\tilde{F}$} \\
    \cline{3-6}
    && $m_a=0\GeV$ & $1\GeV$ & $2\GeV$ & $2.5\GeV$\\
    \hline
    vc-5 & 2.43375(37) & 0.2423(11) & 0.24406(80) & 0.26016(81) & 0.30420(99) \\
    c-5 & 1.94504(23) & 0.22158(58) & 0.2263(12) & 0.23661(57) & 0.26974(47) \\
    c-phys & 1.90218(14) & 0.22125(43) & 0.22283(44) & 0.23703(44) & 0.26960(47) \\
    f-5 & 1.41600(14) & 0.20860(44) & 0.21115(39) & 0.22644(43) & 0.25763(43) \\
    f-phys & 1.37859(13) & 0.20938(41) & 0.21225(37) & 0.22828(42) & 0.26060(49) \\
    sf-5 & 0.93001(16) & 0.20340(67) & 0.20683(68) & 0.22429(76) & 0.25594(93) \\
    \hline
    \hline
  \end{tabular}

\end{table*}

We simultaneously fit the averaged values of $\tilde{C}_{2pt}(T)$ and $C_{2pt}(t)$ on each ensemble to a sum of exponentials representing the tower of vector charmonium states generated by $\mathcal{O}_{J/\psi}$: 
\begin{equation}
\label{eq:fit}
C_{2pt}(t)=\sum_n a_n^2e^{-M_nt}\,\,;\,\,\tilde{C}_{2pt}(T)=\sum_n a_nb_n e^{-M_nT}\;.
\end{equation}
In the expression above we have not shown the oscillatory terms that result from using a staggered quark formalism, but they are included in our fit. The tower of states labelled by $n$ starts at the ground-state, $n=0$, with the $J/\psi$. The higher states are radial and other excitations with the same quantum numbers but larger masses. We are only interested in results for the ground-state here, but it is important to include the other states in the fit to remove excited-state contamination from the ground-state parameters and to make sure that the uncertainties on those parameters from the fit are reliable. We use a standard Bayesian fitting approach~\cite{corrfitter}, including 5 standard and 5 oscillating exponentials and checking that ground-state fit results are stable under a change in the number of exponentials. The $J/\psi$ mass in lattice units, $aM_{J/\psi}$, is the ground-state mass ($M_0$) from the fits using Eq.~\eqref{eq:fit} and the ground-state amplitude, $a_0$, is related to the $J/\psi$ decay constant (although we will not use that here). The ground-state amplitude, $b_0$, is then related to the matrix element for the $J/\psi$ to decay to an on-shell photon and an on-shell ALP in our lattice calculation. From $b_0$ we construct the raw (unnormalised) lattice form factor through
\begin{equation}
\label{eq:rawF}
F_{\mathrm{latt}}(0,m_a)=b_0\frac{\sqrt{2aM_{J/\psi}}}{aq^y} \quad ,
\end{equation}
using Eq.~\eqref{eq:Fdef} and remembering that, as previously noted, we only calculate one of two possible diagrams for the decay with our single 3-point correlation function.  $aq^y$ is  the $y$-component of the photon momentum since our configuration uses $x$ and $z$ polarisations for $J/\psi$ and photon. 

We need to renormalise the currents $j_a$ and $j_{\gamma}$ to match their continuum counterparts in order to determine physical results. 
The renormalisation factor $Z_V$ for the local vector current that we use for $j_{\gamma}$ was determined using the RI-SMOM intermediate scheme in~\cite{Hatton:2019gha,Hatton:2021dvg}. We use the values determined at a scale of $\mu =2$ GeV but note that the small $\mu-$dependence seen is purely a lattice artefact, vanishing in the continuum limit.  To normalise the pseudoscalar current $j_a$ we note that the HISQ action obeys the same partially conserved axial current relation as in the continuum and hence $2am_c^{\mathrm{val}}\overline{c}\gamma_5c$ has the same normalisation as its continuum counterpart (for the $\gamma_5\otimes\gamma_5$ spin-taste pseudoscalar current). Here $am_c^{\mathrm{val}}$ is the bare lattice $c$ mass from Table~\ref{tab:params}. 

The normalised result for $\tilde{F}$ on each set of gluon field configurations is then obtained from 
\begin{eqnarray}
\label{eq:Ftilde-latt}
\tilde{F}_{\mathrm{latt}}&=&b_0\frac{\sqrt{2aM_{J/\psi}}}{aq^y} \frac{2am_c^{\mathrm{val}}}{aM_{J/\psi}} Z_V   \left(1-\frac{m_a^2}{M_{J/\psi}^2}\right)          \quad , \nonumber \\
&=&b_0\frac{\sqrt{2aM_{J/\psi}}}{aq^y} \frac{2am_c^{\mathrm{val}}}{aM_{J/\psi}} Z_V  \frac{2a|\mathbf{q}|}{aM_{J/\psi}}         \quad .
\end{eqnarray}
In the last line we used Eq.~\eqref{eq:momval}.
For our momentum configuration ($q^y=|\mathbf{q}|$, Eq.~\eqref{eq:theta}), this can be simplified to 
\begin{equation}
\label{eq:finalFtilde}
\tilde{F}_{\mathrm{latt}}=Z_Vb_0\sqrt{2aM_{J/\psi}}\frac{4am_c^{\mathrm{val}}}{(aM_{J/\psi})^2} \quad .
\end{equation}

Table~\ref{tab:rawresults} gives the values we obtain for $\tilde{F}_{\mathrm{latt}}$ on each set of gluon field configurations and for each target value of $m_a$. Note the small statistical uncertainties on these values from the fits, typical of calculations for ground-state charmonium mesons from lattice QCD~\cite{Hatton:2020qhk,Colquhoun:2023zbc}. In the next subsection we will discuss how we use these numbers to obtain a curve for $\tilde{F}_{\mathrm{latt}}$ as a function of $m_a$ in the physical continuum limit that can be used to determine constraints on ALP parameters from experimental data. 

\subsection{Determining the physical result for the form factor} \label{sec:continuum}

Our lattice results consist of values for $\tilde{F}$ at 4 different ALP masses at 4 different values of the lattice spacing and 2 different values of the sea $u/d$ quark mass, along with their correlations. We need to convert these into a curve for $\tilde{F}$ as a function of $m_a$ at zero lattice spacing and physical quark masses. This means interpolating between the $m_a$ values and extrapolating to $a=0$ and at the same time adjusting for quark mass mistuning. 
To do this, we fit our lattice data using cubic splines, choosing Steffen splines~\cite{steffen} since they are monotonic between points.\footnote{ A spline $S(X)$ defines a function whose values $S(X_j)$ are specified for each of a discrete set of	`knots' $X_j$. The function is modeled by cubic polynomials between knots  whose coefficients are adjusted so that the function is continuous and has continuous first derivatives at the knots. A monotonic spline, like the Steffen spline, increases or decreases monotonically between knots; maxima and minima can only occur at the knots. The Steffen spline is a particularly robust implementation of a monotonic spline~\cite{steffen}. We used spline software from the \texttt{gvar} Python module in our fits~\cite{gvar}.} We use the following fit form:
\begin{eqnarray}
\label{eq:contfit}
  \tilde{F}_{\mathrm{latt}}(a,X,m_q) &=& S_0(X) + \sum_{i=1}^{i_{\mathrm{max}}}\kappa^{(i)}_{am_c}(am_c)^{2i}S_1^{(i)}(X) \nonumber \\
  &&\hspace{-4.0em}+ \kappa_{\mathrm{val},c} S_2(X)\delta^{\mathrm{val}}_c + \kappa_{\mathrm{sea},c}S_3(X)\delta^{\mathrm{sea}}_c \nonumber \\
  &&\hspace{-5.0em}+ \kappa^{(0)}_{\mathrm{sea},uds}\delta^{\mathrm{sea}}_{uds}\left\{S^{(0)}_4(X)+\kappa^{(1)}_{\mathrm{sea},uds}(am_c)^2S^{(1)}_4(X)\right. \nonumber \\
  &&\hspace{-4.0em}+\left.\kappa^{(2)}_{\mathrm{sea},uds}(am_c)^4S^{(2)}_4(X)\right\},  
\end{eqnarray}
where $X\equiv m_a^2/M_{J/\psi}^2$ ($0<X<1$) and $S_n(X)$ are the cubic splines. The leading spline $S_0(X)$ gives the result for $\tilde{F}$ in the continuum limit with all quark masses fully tuned to their physical values. The value of $X$ associated with each lattice data point is determined from our input value of $\theta$ and our fitted result for $aM_{J/\psi}$ (Table~\ref{tab:rawresults}). By combining Eqs.~\eqref{eq:momval} and~\eqref{eq:theta}, we have 
\begin{equation}
\label{eq:Xval}
X = 1-\frac{2\pi\theta}{aM_{J/\psi}L_s} \, . 
\end{equation}

We include discretisation effects in our fit through the $S^{(i)}_1(X)$ splines, taking the scale of discretisation effects to be the $c$ quark mass (the largest scale present). Only even powers of $a$ appear for the HISQ action~\cite{Follana:2006rc}. Mistuning of sea and valence quarks are accounted for by the $\delta$ terms and the splines $S_2(X)$, $S_3(X)$ and $S^{(k)}_4(X)$. The $\delta$ values are defined as:
\begin{align}
  \delta^{\mathrm{val},c}&=\frac{am_c^{\mathrm{val}}-am_c^{\mathrm{tuned}}}{am_c^{\mathrm{tuned}}},\\
  \delta^{\mathrm{sea},c}&=\frac{am_c^{\mathrm{sea}}-am_c^{\mathrm{tuned}}}{am_c^{\mathrm{tuned}}},\\
  \delta^{\mathrm{sea},uds}&=\frac{2am_l^{\mathrm{sea}}+am_s^{\mathrm{sea}}-2am_l^{\mathrm{tuned}}-am_s^{\mathrm{tuned}}}{10am_s^{\mathrm{tuned}}} \, ,
\end{align}
where the charm quark is tuned so that the $J/\psi$ mass is equal to its experimental value. The valence $c$ quark masses are already very well-tuned but we provide fine adjustments using
\begin{equation}
  am^{\mathrm{tuned}}_c=am^{\mathrm{val}}_c\left(\frac{M^{\mathrm{expt}}_{J/\psi}}{M_{J/\psi}}\right)^{1.5} \, .
\end{equation}
$M^{\mathrm{expt}}_{J/\psi}=3.0969\GeV$ from~\cite{ParticleDataGroup:2024cfk}, and lattice masses $aM_{J/\psi}$ are taken from Table~III in~\cite{Hatton:2020qhk}. The power of 1.5 in the above expression is determined from results in~\cite{Hatton:2020qhk}, and is greater than 1 because of the $J/\psi$ binding energy. We also use the above tuned values to allow for the mistuning of the $c$ sea masses. 

The mass of the $s$ quark is tuned using the (unphysical) $s\overline{s}$ pseudoscalar meson called the $\eta_s$. The `physical' value of the $\eta_s$ mass is $688.5(2.2)\MeV$, determined from lattice QCD in terms of $\pi$ and $K$ masses~\cite{Dowdall:2013rya}. When available we take the values for the tuned $s$ quark masses from Table~V of~\cite{Chakraborty:2014aca}. The only ensemble on which this is not possible is f-5 where we take the $\eta_s$ mass (for the same valence $s$ mass) from~\cite{McLean:2019sds} and use
\begin{equation}
  \label{eq:tune_s}
  am_s^{\mathrm{tuned}}=am_s^{\mathrm{val}}\left(\frac{M_{\eta_s}^{\mathrm{phys}}}{M_{\eta_s}}\right)^2 \, .
\end{equation}
This results in a $\delta^{\mathrm{sea},uds}$ value of $0.0297(17)$ for f-5.

The tuned values for $am_l$  are obtained from the tuned $s$ masses and the ratio~\cite{Bazavov:2017lyh}
\begin{equation}
  \frac{m_s^{\mathrm{phys}}}{m_l^{\mathrm{phys}}}=27.18(10) \, .
  \label{eq:tune_l}
\end{equation}

\begin{figure}
		\includegraphics[width=0.95\linewidth]{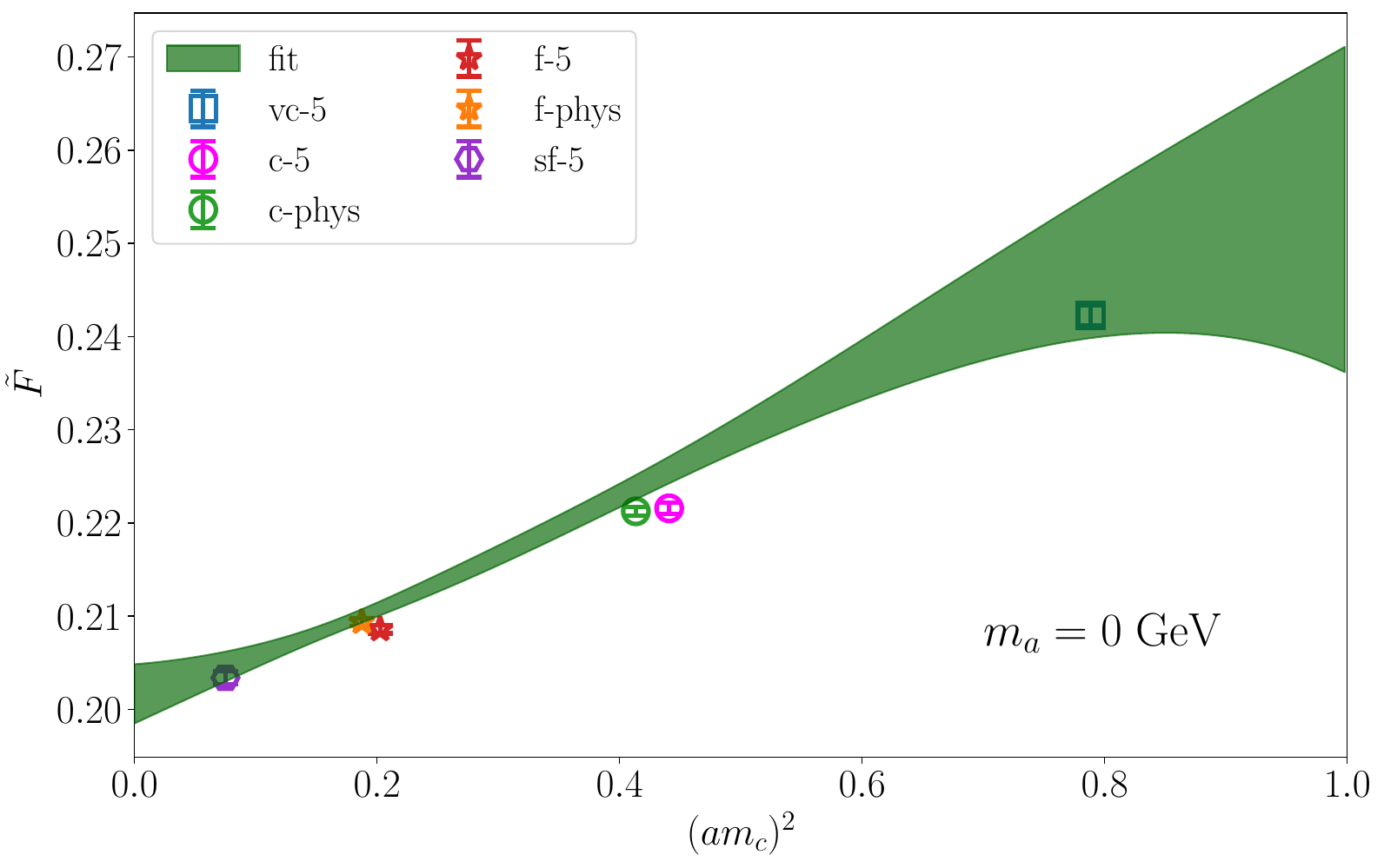}
		\includegraphics[width=0.95\linewidth]{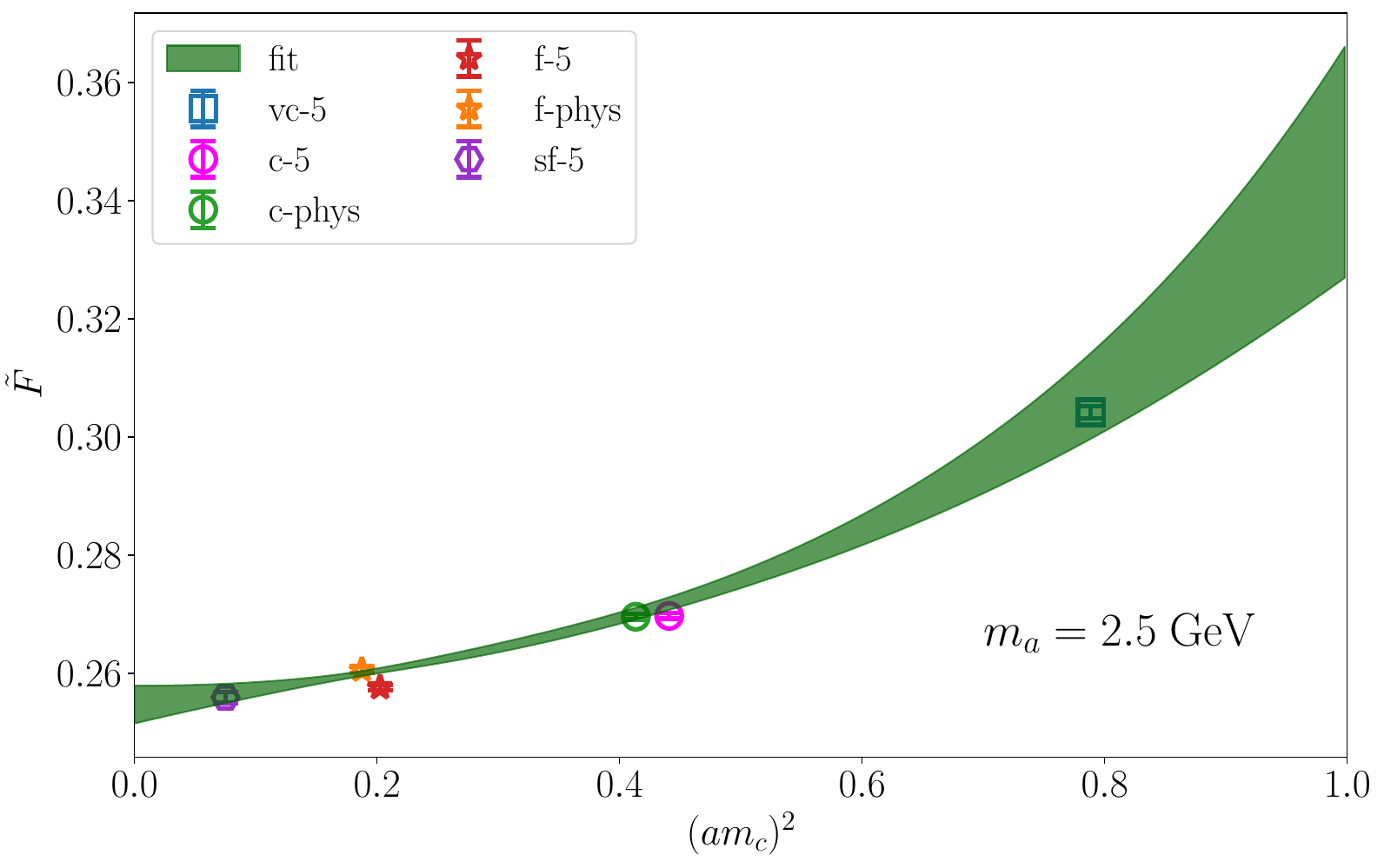}
		\caption{Plots showing the continuum extrapolation of our results for $\tilde{F}$ at two values of $m_a$. At the top, $m_a= 0\,\mathrm{GeV}$ ($X=0$) and below, $m_a= 2.5\,\mathrm{GeV}$ ($X=0.65$). The data points show our lattice results and the green curve is our fit as a function of lattice spacing given in units of the lattice $c$ quark mass. The fit curves shown are the first two terms of Eq.~\eqref{eq:contfit} for the specific values of $X$. }
		\label{fig:disceffects}
\end{figure}

In performing the fit of Eq.~\eqref{eq:contfit}~\cite{lsqfit} we choose two knots for the spline functions just outside our data region at either end. We then test how many additional knots are required by looking at the Bayes Factor. Here the Bayes Factor indicates that a single additional interior knot is optimal. Based on this, we set priors for the position of the knots as $X=-0.22(2)$, $0.40(2)$ and $0.80(2)$. All coefficients $\kappa$ take priors of $0.0(5)$ and spline values at the knots $0.0(3)$. We include discretisation terms up to $(m_ca)^8$, taking $i_{\mathrm{max}}=4$.
The $\chi^2/\mathrm{dof}$ value for our fit is 0.76 for 26 degrees of freedom. 

The final curve for $\tilde{F}$ as a function of $m_a$ at zero lattice spacing will be discussed in Section~\ref{sec:results}. Here we show the dependence of $\tilde{F}$ at two values of $m_a$ as a function of lattice spacing to demonstrate how the discretisation effects are resolved. Figure~\ref{fig:disceffects} shows our results at two ends of our range, $m_a = 0\,\mathrm{GeV}$ and $m_a=2.5\,\mathrm{GeV}$ as a function of $(m_ca)^2$, along with the fit curve  for those values of $m_a$. The fit curve plotted is the first two terms of Eq.~\eqref{eq:contfit} evaluated at an $X$ value equal to the ratio of the squares of our target ALP mass and the experimental $J/\psi$ mass. We see that discretisation effects are relatively mild in both cases, with almost linear dependence on $(m_ca)^2$ giving way to higher orders in $(m_ca)^2$ as $m_a$ is increased. 

\begin{figure}
		\includegraphics[width=0.95\linewidth]{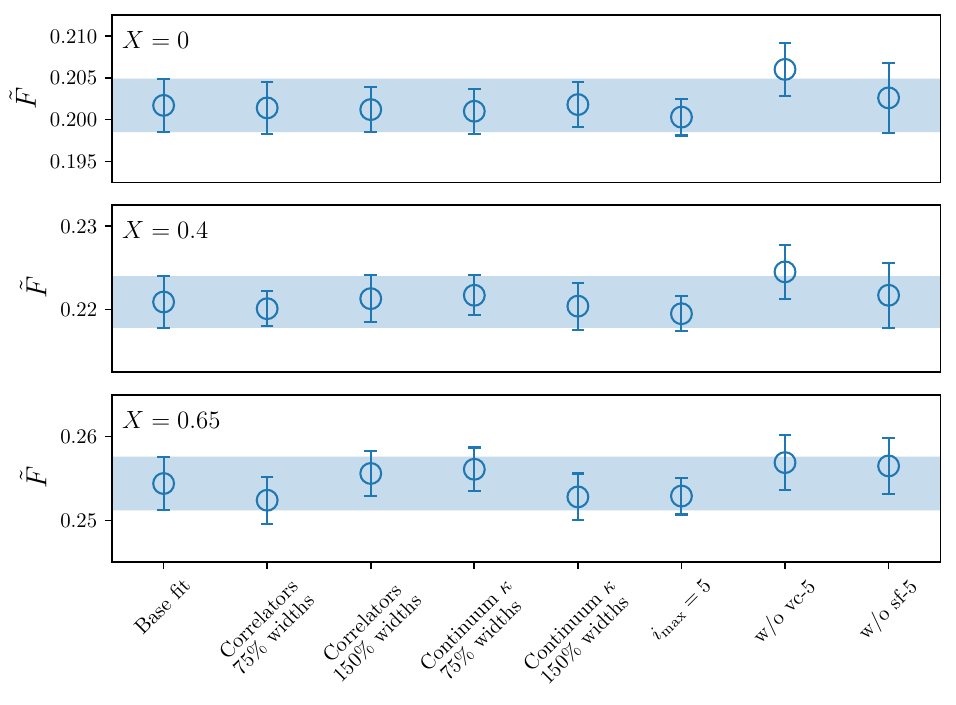}
		\caption{Stability plot for $\tilde{F}$ for three selected values of $m^2_a/M^2_{J/\psi}$ (from top to bottom: $X=0$, 0.4 and 0.65) after adjustments to our base fit shown by the leftmost points and the blue bands. From left to right: we reduce and increase the widths on the priors in the correlator fits, and similarly reduce and increase the prior width on the coefficients $\kappa$ in our spline fit, increase the largest power of $am_c$ included in our discretisation effects from 8 to 10 and drop successively our coarsest and finest gluon field ensemble.}
		\label{fig:stability}
\end{figure}

\begin{figure}
		\includegraphics[width=0.95\linewidth]{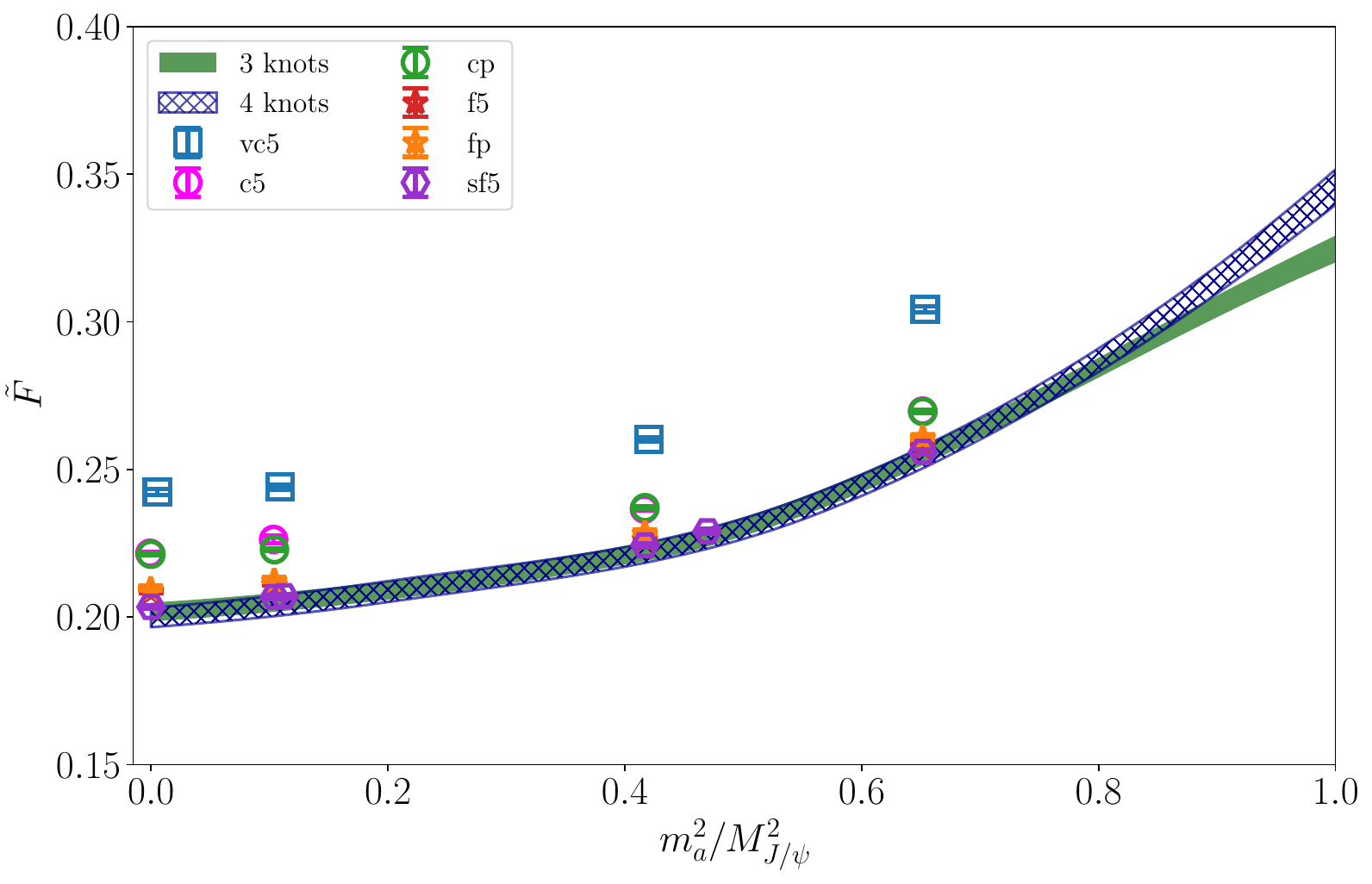}
		\caption{A comparison of 3-knot and 4-knot cubic spline fits for $\tilde{F}$. Our preferred 3-knot fit is given by the green band and the 4-knot fit, with poorer $\chi^2$ and lower Bayes factor, by the blue hashed band. The two fits agree well for $X\equiv m_a^2/M^2_{J/\psi}$ wthin our data region ($0<X<0.65$) and start to pull apart outside that region. The data points are our lattice results on the different sets of gluon field configurations given in Table~\ref{tab:params}.}
		\label{fig:3vs4}
\end{figure}

We close this subsection by showing tests of the robustness of our final result.  
In Figure~\ref{fig:stability} we test the impact on our final results of making changes to the preferred `base' fit we have described above. We test decreasing and increasing the widths of the priors in the fits to the correlators (Eq.~\eqref{eq:fit}), and we similarly change the widths of the priors for the coefficients $\kappa$ in our continuum fit (Eq.~\eqref{eq:contfit}) (keeping the priors on the knot positions and spline values the same). We also change the number of discretisation terms and drop gluon field ensembles at either end of our lattice spacing scale. The variations all give results well within a standard deviation of the base fit for the four illustrated values of $X$ that cover the range over which our data extends ($0<X<0.65$). 

As noted earlier, our preferred fit uses cubic spline functions with 3 knots. In Figure~\ref{fig:3vs4} we show a comparison of our preferred fit with a fit that uses 4 knots. The two interior knots are given priors of 0.20(2) and 0.55(2). The 4-knot fit has a poorer $\chi^2/\mathrm{dof}$ at 1.1 and a considerably lower Bayes factor. The two fits nevertheless agree well for $X$ values within the data range $0<X<0.65$. Above $X=0.65$ they start to pull apart. This illustrates the robustness of our fits for $0<X<0.65$ but also that we must take care in extrapolating results above $X=0.65$. The difference between the 3-knot and 4-knot fits for $X>0.65$ provides a measure for the size of systematic error that can arise from the extrapolation. Since the two fits agree well within 1$\sigma$ up to $X=0.8$ we consider extrapolation to be safe up to this point, but we add an additional (symmetric) systematic error to our $\tilde{F}(X)$ values for $0.65<X<0.8$ that is equal to the size of the difference between the 3-knot and 4-knot fit in that region.

\subsection{Results} \label{sec:results}

\begin{figure}
		\includegraphics[width=0.95\linewidth]{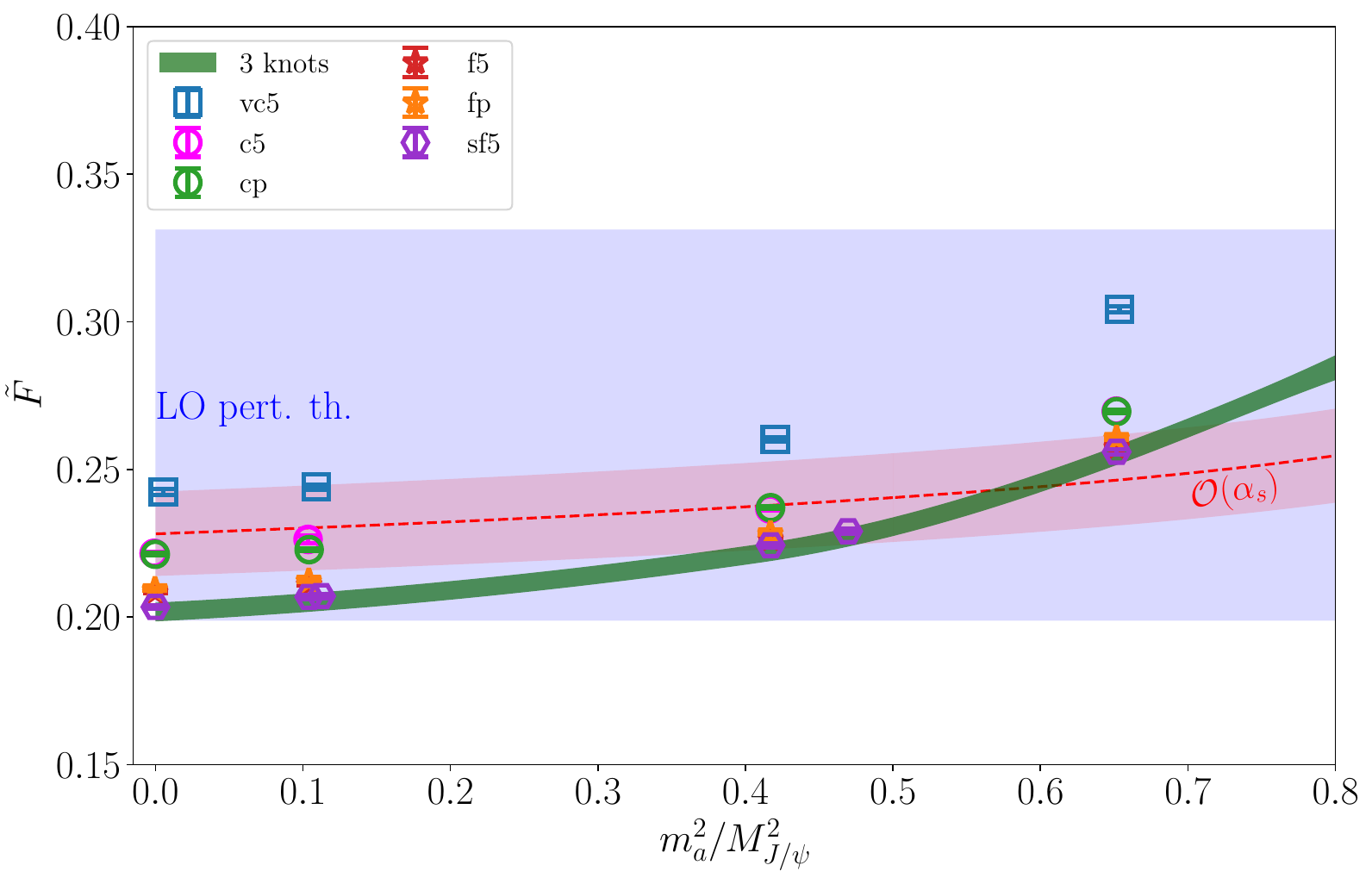}
		\caption{Results for the form factor $\tilde{F}$ plotted against the ratio $m_a^2/M_{J/\psi}^2$. The points show our lattice data for the different sets of gluon field configurations given in Table~\ref{tab:params}. The green band gives our final result in the physical continuum limit from our fit using cubic splines described in the text. The blue dashed line and the corresponding band give the results from leading order NRQCD perturbation theory, while the red dashed line and band includes the first order $\mathcal{O}(\alpha_s)$ corrections (see Eq.~\eqref{eq:Ftildepert2}). The bands represent the uncertainty in the perturbative results, taken as $1\times \alpha_s$ for leading-order and $1\times\alpha_s^2$ for first order, with $\alpha_s=0.25$.}
		\label{fig:formfactor}
\end{figure}

Figure~\ref{fig:formfactor} shows our lattice results for the form factor $\tilde{F}$ (Eq.~\eqref{eq:Ftildedef}) as the data points. The curve, along with its uncertainty, that we obtain for $\tilde{F}$ in the physical continuum limit as a function of the ratio $m_a^2/M_{J/\psi}^2$ is the green band. The data points consist of results at four values of $m_a$ (six values for our finest lattices) on six different sets of gluon field configurations with different values of the lattice spacing and sea $u/d$ quark masses. The data points show smooth monotonic curves for each set with the values of $\tilde{F}$ falling slightly on the finer lattices. By interpolating between the points using cubic splines and allowing for discretisation and quark-mass mistuning effects as described in subsection~\ref{sec:continuum} we obtain the green curve at zero lattice spacing and physical quark masses. The green curve takes much the same shape as the results on the individual sets of gluon fields but lies below them because of the extrapolation to zero lattice spacing. We note that the green curve extends somewhat above our data region (up to $X\equiv m_a^2/M^2_{J/\psi}=0.8$) and we include an additional systematic uncertainty in the green band for $0.65<X<0.8$ to allow for the uncertainty associated with the extrapolation, as described in Section~\ref{sec:continuum}.

Figure~\ref{fig:formfactor} also shows a comparison to the results from leading-order and next-to-leading-order nonrelativistic perturbation theory from Eq.~\eqref{eq:Ftildepert2}. The leading-order result is given by the blue dashed line with an uncertainty band of $\pm25\%$ for missing $\mathcal{O}(\alpha_s)$ effects. Besides this large uncertainty, which does encompass the full result from the lattice QCD calculation, we see that the leading-order perturbative result is completely flat and so does not have the correct shape as $m_a$ varies. 
The red dashed line and band gives the first-order result, now including $\mathcal{O}(\alpha_s)$ corrections with coefficient $b_P$ which depends on $m_a$. We see that the perturbative curve now has a better shape, more in line with the lattice QCD result. The values of $\tilde{F}$ are not in very good agreement with the green curve, however, even allowing for the uncertainty given by the red band. This uncertainty was taken as $1\times[\alpha_s(M_{J/\psi})]^2$, and the disagreement indicates sizeable higher-order corrections are present. Some of the disagreement could be down to relativistic corrections missing from the perturbative calculation, as discussed in Section~\ref{sec:pert}.  Going forward, the lattice QCD result should be used to set ALP constraints rather than the perturbative curves because it incorporates QCD and relativity effects fully. 

Note that we have truncated the $x$-axis of Figure~\ref{fig:formfactor} at $X=0.8$. As discussed in Section~\ref{sec:lattice} there is a practical limit to how high $X$ can go in the lattice QCD calculation. There is also a limit in principle on all calculations coming from neglected mixing between the ALP and the $\eta_c$ as their masses approach each other. $X=0.8$ corresponds to $m_a=$ 2.770 GeV, which is 214 MeV, or $7\Gamma(\eta_c)$, below the $\eta_c$ mass. It should be safe to neglect mixing up to this $X$ value. 

\begin{table}
  \caption{Upper table: Results for $\tilde{F}$ in the physical continuum limit at selected values of $X=m_a^2/M_{J/\psi}^2$. Lower table: correlation matrix for the values in the upper table. }
  \label{tab:Fcontvals}
  \begin{tabular}{cc}
    \hline
    \hline
    $m^2_a/M^2_{J/\psi}$ & $\tilde{F}$\\
    \hline
    0 & 0.2017(32)    \\
    0.4 & 0.2209(31)  \\
    0.65 & 0.2544(32) \\
    \hline
    \hline
  \end{tabular}
\\
\vspace{1.0em}
    \begin{tabular}{c|ccc}
    \hline
    \hline
  & $\tilde{F}(0)$ & $\tilde{F}(0.4)$ & $\tilde{F}(0.65)$  \\
\hline
$\tilde{F}(0)$ & 1.000 & 0.962 & 0.909  \\
$\tilde{F}(0.4)$ & 0.962 & 1.000 & 0.982 \\
$\tilde{F}(0.65)$ & 0.909 & 0.982 & 1.000  \\
\hline
\hline
  \end{tabular}
\end{table}

We give numerical results for $\tilde{F}$ for a choice of three values of $m_a^2/M_{J/\psi}^2$ covering our data region. The values for $\tilde{F}$ are correlated at different values of $m_a$ and we also give the correlation matrix for these values. The correlation matrix is not normally needed to set ALP constraints from experimental data but we include it for completeness. 
 
Table~\ref{tab:errorbudget} gives the error budget for $\tilde{F}$ at each of these $X$ values. The largest sources of uncertainty are from statistics and the extrapolation to the physical point. The total uncertainty is less than 2\% in all cases. There are additional systematic uncertainties from quark-line disconnected diagrams and the effect of electric charge of the $c$ quarks, both of which are missing from our calculation (as well as from the earlier perturbative calculations). We expect both of these uncertainties to be negligible. The total impact of quark-line disconnected diagrams on the $J/\psi$ can be judged from its very small decay width via 3 gluons to hadrons. Quark-line disconnected diagrams allow the mixing of the $J/\psi$ with other flavour-singlet vector mesons, such as the $\phi$, that could then decay to $\gamma a$ and this would be a systematic error here. The very small probability with which this happens is indicated by the branching fraction for $J/\psi$ decay to $K\overline{K}$ (the channel dominating $\phi$ decay), which is less than 0.1\%. Light meson decay to $\gamma a$ is also expected to be suppressed by the appearance of the quark mass in the ALP-quark coupling in Eq.~\eqref{eq:L-int-rewrite}. We therefore expect the systematic uncertainty in $\tilde{F}$ from missing quark-line disconnected diagrams to be well below 0.1\%. The impact of the electric charge of the $c$ quark on the internal structure of the $J/\psi$ meson can be assessed from the effects measured in lattice QCD+QED for the $J/\psi$ decay constant in Ref.~\cite{Hatton:2020qhk}. There a 0.2\% change in $f_{J/\psi}$ was seen, which would translate to a 0.2\% effect in $\tilde{F}$ using the tree-level perturbative formula (Eq.~\eqref{eq:Ftildepert2}). This is also a negligible uncertainty. 

As supplementary material we provide a text file containing our lattice QCD results for $\tilde{F}$ as a function of $X=m_a^2/M_{J/\psi}^2$ covering the range from 0.0 to 0.8 in increments of 0.01. The total uncertainty on the value of $\tilde{F}$ is also given. 

\begin{table}
  \caption{Error budget for $\tilde{F}$ from our fit of the lattice data to cubic splines. The errors are given as a percentage of the final answer. The uncertainty labelled `spline priors' is from priors on $S_0$ values and the knot positions. For a discussion of the negligible additional systematic uncertainties from missing quark-line disconnected diagrams and QED effects, see text. }
  \label{tab:errorbudget}
  \begin{tabular}{c|cccc}
    \hline
    \hline
      & $\tilde{F}(0)$ & $\tilde{F}(0.4)$ & $\tilde{F}(0.65)$  \\
    \hline
    Statistics        & 0.68 & 0.58 & 0.56  \\
    $a^2\to 0$        & 1.26 & 1.14 & 0.98  \\
    Valence mistuning & 0.00 & 0.00 & 0.00  \\
    Sea mistuning     & 0.67 & 0.62 & 0.55  \\
    Spline priors    & 0.04 & 0.12 & 0.07  \\
    \hline
    Total (\%) & 1.58 & 1.43 & 1.26  \\
    \hline
    \hline
  \end{tabular}
\end{table}

\section{ALP constraints using lattice QCD results}\label{sec:constraints}

The process $J/\psi\to \gamma a$ has been searched for at collider experiments, most recently at BESIII~\cite{BESIII:2022rzz,BESIII:2021ges,BESIII:2020sdo}. In this section we study the resulting constraints on ALP parameter space, highlighting differences in constraints calculated using our new lattice QCD result with those calculated with perturbation theory.

\begin{figure}
\begin{center}
\includegraphics[width=0.4\textwidth]{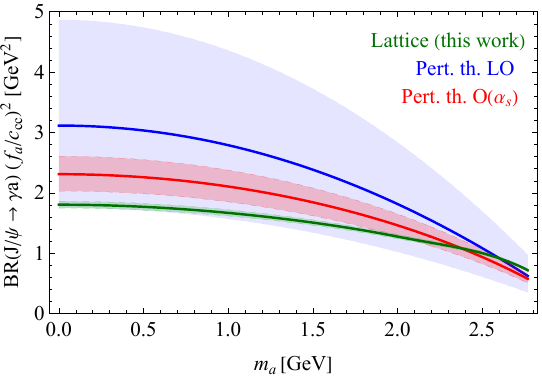}
\caption{\label{fig:BR}Branching ratio for the process $J/\psi \to \gamma a$ for an ALP coupled only to charm quarks, normalised by $(f_a/c_{cc})^2$. The lattice QCD result based on the calculations presented here in Fig.~\ref{fig:formfactor} is shown in green, while results calculated in the perturbative approach at leading order (LO) and at $O(\alpha_s)$ are shown in blue and red respectively.}
\end{center}
\end{figure}

For an ALP which couples only to charm quarks, the branching ratio of $J/\psi\to \gamma a$ is shown in Fig.~\ref{fig:BR}, normalised by the coupling $c_{cc}(\mu_c)$ and the ALP decay constant $f_a$. 
This has been obtained by evaluating Eq.~\eqref{eq:full-rate} with $c_{\gamma\gamma}=0$ and dividing by the experimental total width of the $J/\psi$~\cite{ParticleDataGroup:2024cfk}. We use $\alpha = 1/134.02$, appropriate to charmonium scales in the on-shell scheme~\cite{Hatton:2020qhk}. The branching ratios from perturbation theory use Eq.~\eqref{eq:Ftildepert2} taking $f_{J/\psi}$ from lattice QCD~\cite{Hatton:2020qhk}. The errors on these results have been estimated by applying a multiplicative error of $\alpha_s(m_{J/\psi})=0.25$ to the leading order amplitude, and of $[\alpha_s(m_{J/\psi})]^2=0.06$ to the $O(\alpha_s)$ amplitude. We see that, for most of the ALP mass range, the branching ratio from lattice QCD is below those obtained in perturbation theory, i.e. the perturbative expressions overestimate the branching ratio. 

\subsection{Couplings and decay modes of the ALP}

The sensitivity of different searches in radiative $J/\psi$ decays on ALP parameter space depends on the decay mode and lifetime of the ALP. In general, the ALP can decay to a charged lepton pair, a pair of photons, or hadrons (if $m_a>3 m_\pi$). 
The widths for ALP decay to leptons and photons are
\begin{equation}
\Gamma(a\to l^+l^-)= \frac{m_a m_l^2}{8\pi f_a^2} c_{ll}^2(\mu_a)\sqrt{1-\frac{4m_l^2}{m_a^2}}
\end{equation}
and 
\begin{equation}
\label{eq:atogamgam}
\Gamma(a\to \gamma\gamma)= \frac{\alpha^2 m_a^3}{64\pi^3 f_a^2} \left|c_{\gamma\gamma}^{\text{eff}}(\mu_a)\right|^2 \, .
\end{equation}
For light ALPs below the $3\pi$ threshold these two decay modes dominate. For ALPs with masses above 1.5 GeV but below the $\bar c c$ threshold, decay to light hadrons dominates; the inclusive hadronic decay rate is given by~\cite{Spira:1995rr,Bauer:2021mvw}:
\begin{align}
\Gamma(a\to \text{had.})&=\frac{\alpha_s^2(\mu_a)m_a^3}{8\pi^3f_a}\left[1+\frac{83}{4}\frac{\alpha_s(\mu_a)}{\pi} \right]\left|c_{GG}^{\text{eff}}(\mu_a)\right|^2. \label{eq:decaytohadrons}
\end{align}

We make simple assumptions on the couplings of the ALP at the high scale $\Lambda\equiv 4\pi f_a$, and calculate its branching ratios and lifetimes by evolving these couplings down to $\mu_a$ using their renormalisation group equations (RGEs)~\cite{Choi:2017gpf,Chala:2020wvs,Bauer:2020jbp}. 

The couplings of the ALP to gauge bosons are scale independent~\cite{Chetyrkin:1998mw}. However, the $a\to \gamma\gamma$ and $a\to GG$ decay rates receive important contributions from finite loop graphs involving light fermions and gluons. These can be accounted for by defining an effective coupling $c_{\gamma\gamma}^{\mathrm{eff}}$ in the decay rate of Eq.~\eqref{eq:atogamgam}, and an effective coupling $c_{GG}^{\text{eff}}$ in the decay rate of Eq.~\eqref{eq:decaytohadrons}. If the ALP has a mass well above the QCD scale, then these are calculated as~\cite{Bauer:2017ris}
\begin{align}
\label{eq:cgamgamhighmass}
c_{\gamma\gamma}^{\mathrm{eff}}(\mu_a)&=c_{\gamma\gamma}+\sum_{f\neq t}N_c^fQ_f^2c_{ff}(\mu_a) B_1(\tau_f), \\
c_{GG}^{\mathrm{eff}}(\mu_a)&=c_{GG}+\frac{1}{2}\sum_{q\neq t}c_{qq}(\mu_a) B_1(\tau_q),
\end{align}
where $N_c^q=3$ for quarks, and $N_c^l=1$ for leptons, $Q_f$ is the electric charge of fermion $f$, and $\tau_f=4m_f^2/m_a^2$, with
\begin{equation}
B_1(\tau)=1-\tau f^2(\tau), ~~f(\tau)=\begin{cases} \arcsin\frac{1}{\sqrt{\tau}} & \tau \geq 1, \\ \frac{\pi}{2}+\frac{i}{2}\ln \frac{1+\sqrt{1-\tau}}{1-\sqrt{1-\tau}} & \tau < 1. 
\end{cases}
\end{equation}
This function behaves as $B_1(\tau_f)\approx 1$ for $m_f \ll m_a$, and $B_1(\tau_f)\approx -\frac{m_a^2}{12m_f^2}$ for $m_f \gg m_a$, meaning that charged fermions \emph{lighter} than the ALP contribute significantly to $c_{\gamma\gamma}^{\mathrm{eff}}$ (and $c_{GG}^{\mathrm{eff}}$), but charged fermions \emph{heavier} than the ALP decouple. For ALP masses below the QCD scale, the light quark contributions to $c_{\gamma\gamma}^{\mathrm{eff}}$ can be calculated using chiral perturbation theory as~\cite{Bauer:2017ris}
\begin{equation}
c_{\gamma\gamma}^{\mathrm{eff}}(\mu_a)=c_{\gamma\gamma}-\frac{1}{2}\frac{m_a^2}{m_\pi^2-m_a^2}\left(c_{uu}(\mu_a)-c_{dd}(\mu_a)\right) + \ldots
\end{equation}
where omitted terms include contributions from ALP-gluon couplings, and contributions from leptons and heavy quarks, which are as in Eq.~\eqref{eq:cgamgamhighmass}. The phenomenological implication of the mass dependence of $c_{\gamma\gamma}^{\mathrm{eff}} (\mu_a)$ is that an ALP produced in $J/\psi$ decays with flavour-universal couplings to up-type quarks can decay to photons via its up-quark coupling (assuming $m_a>m_u$). On the other hand, if an ALP with mass $m_a< m_c$ couples only to charm quarks, its decay rate to photons will be very small, and it is likely to be long-lived.

The ALP-fermion couplings are scale-dependent, and their one-loop RGEs depend in general on all other Wilson coefficients in the dimension-five effective Lagrangian. However, many terms are suppressed by small Yukawa couplings. Neglecting all Yukawas except $y_t$, the RGE (above the electroweak scale) for the ALP-lepton couplings is (for $l=e,\mu,
\tau$)~\cite{Chala:2020wvs,Bauer:2020jbp}:
\begin{equation}
\label{eq:RGEcll}
\frac{d}{d\ln \mu}c_{ll}(\mu)=\frac{3}{16\pi^2}\bigg(-2y_t^2 c_{tt}(\mu)+3\alpha_2^2 c_{WW}+5\alpha_1^2c_{BB} \bigg),
\end{equation}
where $c_{WW}$ and $c_{BB}$ are the Wilson coefficients multiplying the Lagrangian terms $\alpha_2a W_{\mu\nu}^A\tilde{W}^{\mu\nu,A}/(4\pi f_a)$ and  $\alpha_1a B_{\mu\nu}\tilde{B}^{\mu\nu}/(4\pi f_a) $ respectively, which couple the ALP to the electroweak gauge bosons\footnote{The coefficient $c_{\gamma\gamma}$ is expressed in terms of $c_{WW}$ and $c_{BB}$ as $c_{\gamma\gamma}=c_{WW}+c_{BB}$~\cite{Bauer:2020jbp}.}. The presence of the term proportional to $c_{tt}(\mu)$ means that an ALP with flavour-universal couplings to up-type quarks at $\Lambda$ will inherit a sizeable coupling to charged leptons at lower scales, allowing its decay to electrons and muons. Specifically, the numerical solution to the RGE for flavour-diagonal couplings when running  from $\Lambda=4\pi\,$TeV to $\mu_0=2\,$GeV is~\cite{Bauer:2021mvw}:
\begin{equation}
c_{ll}(\mu_0)\approx c_{ll}(\Lambda) + 0.12\,c_{tt}(\Lambda)+ \ldots
\label{eq:RGEcllsol}
\end{equation}
where additional terms denoted by `$\ldots$' are proportional to the gauge boson couplings and the other fermionic couplings, with prefactors of $O(10^{-4})$ or smaller.

To calculate the branching ratio of $J/\psi\to \gamma a$, it is important to also understand the running of the ALP-charm couplings. The form of this is very similar to that for the ALP-lepton couplings~\eqref{eq:RGEcll}, but with different prefactors corresponding to its different electroweak couplings and colour charge~\cite{Chala:2020wvs,Bauer:2020jbp}:
\begin{eqnarray}
\frac{d}{d\ln \mu}c_{cc}(\mu)&=&\frac{3}{16\pi^2}\bigg(2y_t^2 c_{tt}(\mu)+\frac{32}{3}\alpha_s^2 c_{GG}\\&&\hspace{3.0em}+3\alpha_2^2 c_{WW} +\frac{17}{9}\alpha_1^2c_{BB}\bigg).\nonumber \label{eq:RGEccc}
\end{eqnarray}
The numerical solution to this RGE when running  from $\Lambda=4\pi\,$TeV to $\mu_0=2\,$GeV is~\cite{Bauer:2021mvw}:
\begin{equation}
c_{cc}(\mu_0)\approx c_{cc}(\Lambda) - 0.12\,c_{tt}(\Lambda)-0.04\, c_{GG}+\ldots \label{eq:RGEcccsol}
\end{equation}
where additional terms denoted by `$\ldots$' are proportional to the electroweak gauge boson couplings and the other fermionic couplings, with prefactors of $O(10^{-4})$ or smaller. Similar to the case of the leptonic couplings, we see that the top coupling $c_{tt}$ has the largest impact on the running of $c_{cc}$.

\subsection{Confronting experimental searches}

\begin{figure}
\begin{center}
\includegraphics[width=0.35\textwidth]{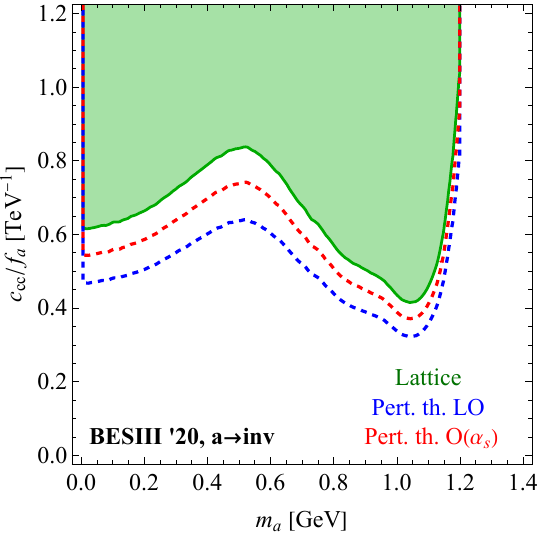}
\caption{\label{fig:BESIII20cccvsmass}Constraints on ALP parameter space from a BESIII search \cite{BESIII:2020sdo} for an invisible particle produced in radiative $J/\psi$ decays, assuming that the ALP has only couplings to charm quarks. In this case, the ALP is naturally long-lived across all relevant parameter space (with proper decay lengths of at least several metres). The green region is excluded by the lattice QCD results reported here; the blue and red dashed lines give the central value of the boundary of the excluded region using the earlier perturbative calculation. }
\end{center}
\end{figure}

The BESIII search of Ref.~\cite{BESIII:2020sdo} looked for $J/\psi \to \gamma +$invisible, making it sensitive to ALPs which are stable on detector scales and in the mass range $m_a<1.2$ GeV. This is the case for an ALP which couples only to charm quarks, since its loop-induced photon decay (see Eq.~\eqref{eq:cgamgamhighmass}) is suppressed to the extent that it is long-lived across the parameter space relevant for the search. The coupling $c_{cc}$ is also essentially scale-invariant in this scenario since its self renormalisation is proportional to the square of the small charm Yukawa (see Eq.~\eqref{eq:RGEcccsol}). 
Upper limits on $c_{cc}$  can therefore be found by comparing the theoretical branching ratio (Eq.~\eqref{eq:full-rate} with $c_{\gamma\gamma}=0$ and Fig.~\ref{fig:BR}) to the experimental (90\% confidence level) upper limit for the $J/\psi$ branching fraction at each value of ALP mass. 
The resulting excluded region for $c_{cc}$ shown in green in Fig.~\ref{fig:BESIII20cccvsmass} is calculated using the lattice result in this work, while those in dotted blue and red are calculated using the central values of the perturbative calculation at leading order and to $O(\alpha_s)$, respectively. It can be seen that using the perturbative result leads to a significant overestimate of the sensitivity, especially in the leading order case.

\begin{figure}
\begin{center}
\includegraphics[width=0.35\textwidth]{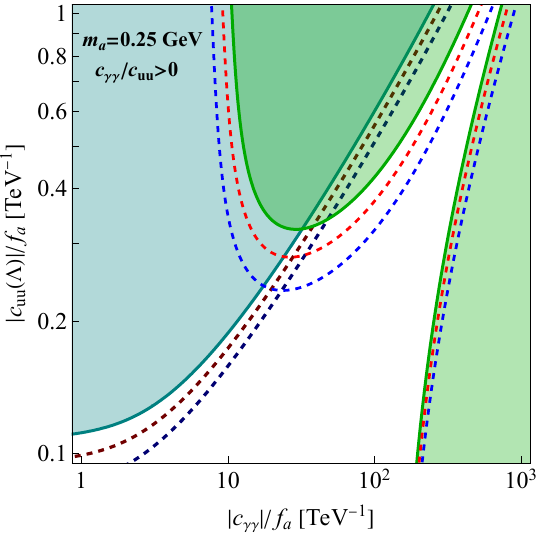}
\caption{ \label{fig:mumugamgampt25gev}  Constraints from BESIII searches for $J/\psi \to \gamma a$ assuming an ALP mass of 0.25 GeV, with the ALP decaying into $\mu^+\mu^-$ \cite{BESIII:2021ges} (in teal) and with the ALP decaying into $\gamma\gamma$ \cite{BESIII:2022rzz} (in green). Constraints are shown on the absolute value of the couplings, under the assumption that they have the \emph{same sign}, i.e.~$c_{\gamma\gamma}/c_{uu}(\Lambda)>0$. The solid teal and green exclusion regions were found using our new lattice QCD results for $\tilde{F}$. Corresponding bounds found using the central value of the LO (blue) and $O(\alpha_s)$ (red) perturbation theory calculations are shown with dotted lines. The coupling of the ALP is assumed to be \emph{flavour universal} to up-type quarks, i.e.~$c_{uu}(\Lambda)=c_{cc}(\Lambda)=c_{tt}(\Lambda)$. }
\end{center}
\end{figure}
\begin{figure}
\begin{center}
\includegraphics[width=0.35\textwidth]{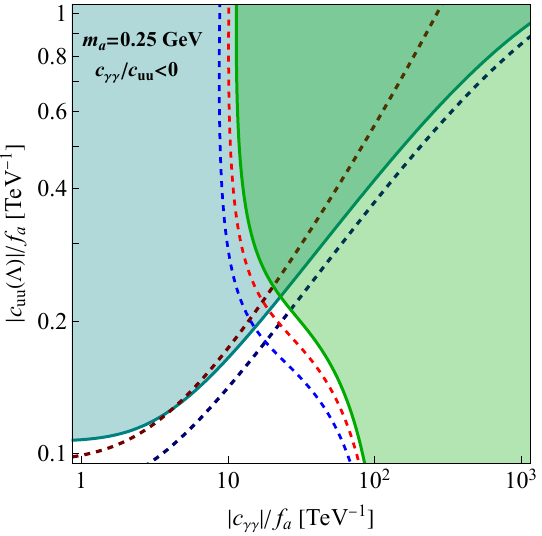}
\caption{ \label{fig:mumugamgampt25gevneg} As in Fig.~\ref{fig:mumugamgampt25gev}, but now taking the couplings to be of \emph{opposite} signs; $c_{\gamma\gamma}/c_{uu}(\Lambda)<0$. }
\end{center}
\end{figure}

\begin{figure}
\begin{center}
\includegraphics[width=0.35\textwidth]{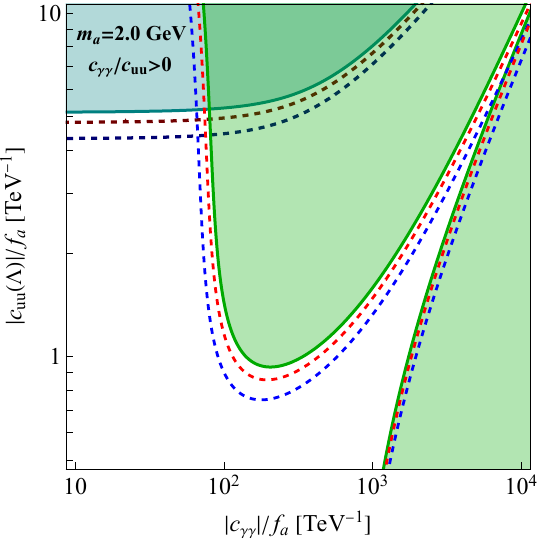}
\caption{ \label{fig:mumugamgam2gev} As in Fig.~\ref{fig:mumugamgampt25gev}, but for an ALP mass of $m_a=2$ GeV. Couplings are assumed to be of the same sign; $c_{\gamma\gamma}/c_{uu}(\Lambda)>0$.}
\end{center}
\end{figure}
\begin{figure}
\begin{center}
\includegraphics[width=0.35\textwidth]{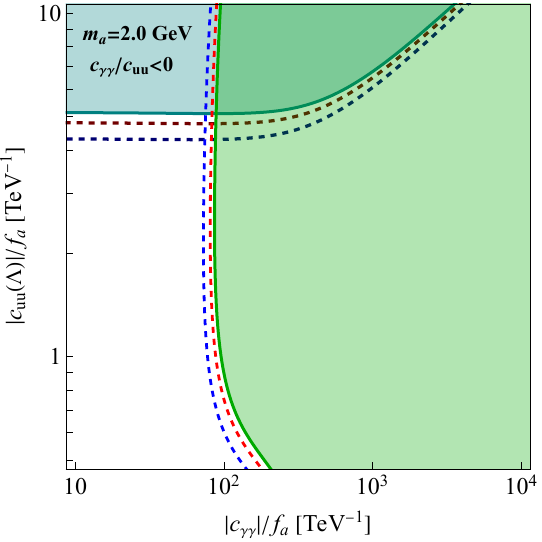}
\caption{ \label{fig:mumugamgam2gevneg} As in Fig.~\ref{fig:mumugamgam2gev}, but now taking the couplings to be of \emph{opposite} signs; $c_{\gamma\gamma}/c_{uu}(\Lambda)<0$.}
\end{center}
\end{figure}

An arguably more realistic ALP model has flavour universal couplings to up-type quarks at scale $\Lambda$, along with tree level couplings to photons, but with couplings to down-type quarks and leptons set to zero at $\Lambda$. In this case, the $J/\psi \to \gamma a$ decay proceeds through both diagrams in Fig.~\ref{fig:feyn-diags}.
Through renormalisation group running, Eq.~\eqref{eq:RGEcll}, an ALP coupling to top quarks at the UV scale $\Lambda$ will generate a sizeable coupling to charged leptons at the ALP mass scale, allowing $a\to l^+l^-$ decays. Such an ALP can therefore be constrained by the BESIII search for a pseudoscalar particle produced in radiative $J/\psi$ decays and decaying to a pair of muons \cite{BESIII:2021ges}. If the ALP also has couplings to photons, then the BESIII search for an ALP decaying to a photon pair in the $J/\psi\to \gamma\gamma\gamma$ process \cite{BESIII:2022rzz} provides additional constraints. We set the low scale for the couplings as $\mu_a = \mu_c = 2\,\mathrm{GeV}$, and solve for the couplings at the low scale in terms of the couplings at the scale $\Lambda$, which we take to be fixed at $\Lambda=4\pi$ TeV. The solutions to the RGEs for the leptonic and charm couplings are then as given in Eq.~\eqref{eq:RGEcllsol} and Eq.~\eqref{eq:RGEcccsol} (with $c_{GG}=0$). This enables us to convert the experimental 90\% C.L.\@ upper limits on the branching fractions into upper limits on a combination of $c_{cc}(\Lambda)$ and $c_{\gamma\gamma}$. 

The resulting limits in the 2-d plane of these couplings using our lattice results for $\tilde{F}$ are shown in Figs.~\ref{fig:mumugamgampt25gev} and~\ref{fig:mumugamgampt25gevneg}, for an ALP mass of $m_a=0.25$ GeV (below the $3\pi$ threshold). In teal is the excluded region from the search in the muonic final state~\cite{BESIII:2021ges}, while the green region is excluded by the search for an ALP decaying to photons \cite{BESIII:2022rzz}. The quark coupling plotted on the vertical axis is assumed to be \emph{flavour universal}, i.e.~$c_{uu}(\Lambda)=c_{cc}(\Lambda)=c_{tt}(\Lambda)$. 
The difference between Fig.~\ref{fig:mumugamgampt25gev} and Fig.~\ref{fig:mumugamgampt25gevneg} is only in the relative sign of the couplings. As can be seen in the width formula, Eq.~\eqref{eq:full-rate}, if the charm coupling and the photon coupling are of the same sign, their contributions can cancel in the rate, which is the reason for the unconstrained area which extends towards the top right of the plot in Fig.~\ref{fig:mumugamgampt25gev}. No such cancellation can occur if the couplings have opposite signs, hence the different form of the excluded regions in Fig.~\ref{fig:mumugamgampt25gevneg}.
Excluded regions calculated using the perturbative approach are shown on these plots with dotted lines; blue lines correspond to the central value of the leading order calculation, while red lines correspond to the central value of the $O(\alpha_s)$ calculation. In some regions of parameter space it can be seen that there is a significant difference between the lattice and perturbative approaches. As expected, this difference is smaller in regions where the quark couplings are smaller and/or the photon coupling is larger, since in this case the $J/\psi\to \gamma a$ decay is proceeding predominantly through the right-hand diagram in Fig.~\ref{fig:feyn-diags}, which is unaffected by the new lattice calculation of $\tilde{F}$.

Figures~\ref{fig:mumugamgam2gev} and~\ref{fig:mumugamgam2gevneg} shows the same constraints as in Figs.~\ref{fig:mumugamgampt25gev} and \ref{fig:mumugamgampt25gevneg}, but now for an ALP mass of $m_a=2$ GeV. At this heavier ALP mass, the constraints from both searches are not as strong in the $c_{\gamma\gamma}-c_{uu}$ parameter space as they were for a lighter ALP mass. This is mostly due to the fact that a heavier ALP can decay at tree level to hadrons through its quark couplings (see Eq.~\eqref{eq:decaytohadrons}), which significantly reduces its branching ratios to photons and muons. It can also be seen that the differences between the constraints found with the lattice calculation and those found with the perturbative calculations are less significant at this mass range, as would be expected from the fact that the central values of the respective branching ratio calculations are closer at intermediate ALP masses than at low ALP masses, as seen in Fig.~\ref{fig:BR}. The different overall shapes of the constraints in Fig.~\ref{fig:mumugamgam2gev} and Fig.~\ref{fig:mumugamgam2gevneg} are again due to the different assumptions on the relative signs of the couplings; if $c_{cc}$ and $c_{\gamma\gamma}$ have the same sign, their effects can (partially) cancel in $\Gamma(J/\psi \to \gamma a)$, leading to the unconstrained area extending to large couplings towards the top right of Fig.~\ref{fig:mumugamgam2gev}.

\section{Conclusions} \label{sec:conclusions}

The calculation presented here is the first lattice QCD determination of the form factor for $J/\psi \to \gamma a$, allowing for experimental searches for ALPs that couple to $c$ quarks from $J/\psi$ radiative decay, for example by BESIII.  The result we obtain for the quantity $\tilde{F}$, defined in Eq.~\eqref{eq:Ftildedef}, is much more accurate than the perturbative calculations used previously. We see from the comparison plot given in Fig.~\ref{fig:formfactor} that $\tilde{F}$ shows mild dependence on the ALP mass and this is missing entirely in tree-level perturbation theory and only partially generated at $\mathcal{O}(\alpha_s)$. The $J/\psi$ branching ratios for an ALP coupling only to $c$ quarks are compared in Fig.~\ref{fig:BR}. 

In future the results for $\tilde{F}$ from lattice QCD along with Eq.~\eqref{eq:full-rate} should be used to set ALP constraints from $J/\psi \to \gamma a$. For this purpose our lattice QCD $\tilde{F}$ values are provided in a text file, HPQCD\_Ftilde.txt, given in the Supplementary Materials. 

Here we show the constraints on ALP masses and couplings using the lattice QCD results along with experimental data from BESIII in Figs.~\ref{fig:BESIII20cccvsmass},~\ref{fig:mumugamgampt25gev},~\ref{fig:mumugamgampt25gevneg},~\ref{fig:mumugamgam2gev} and~\ref{fig:mumugamgam2gevneg}. These figures show that the previously used perturbative approach to the theoretical calculation of the rate overestimates the constraints obtained for almost all of the ALP mass range covered. 

This calculation opens the way for further use of lattice QCD for form factors needed for ALP searches. One obvious extension is to calculate the form factors for $\Upsilon \to \gamma a$ which covers a somewhat larger range of ALP masses than $J/\psi$ decay. This could be done using the techniques demonstrated here and working at $b$ quark masses on very fine lattices with the HISQ formalism as used for the study of $\eta_b \to \gamma\gamma$ decay in~\cite{Colquhoun:2024wsj}. 

Further possibilities include charged pseudoscalar leptonic decay with ALP radiation, such as $B\to \ell \overline{\nu} a$ or $D_s \to \ell \overline{\nu} a$. The calculation of the form factor for this in lattice QCD would proceed in the same way as here, but now with a range of $q^2$ values for the virtual $W$ for each ALP mass. An additional complication is that there are separate form factors for emission of the ALP from the quark and antiquark of different flavour. The computational cost is then substantially larger than that for $J/\psi \to \gamma a$ but clearly tractable. The results would provide complementary information on ALP-quark couplings to that from other processes. Currently approximate methods are being used to determine the relevant form factors~\cite{Gallo:2021ame} and, as here, constraints would be made significantly more compelling with accurate lattice QCD results. In addition this would provide the incentive for improved experimental searches. 

\section*{Acknowledgements} \label{sec:ack}
We thank the MILC collaboration for making publicly available their gauge configurations and their code MILC-7.7.11~\cite{MILCgithub}. 
This work used the DiRAC Data Intensive Service (CSD3) at the University of Cambridge, managed by the University of Cambridge Information Services on behalf of the Science and Technology Facilities Council (STFC) DiRAC HPC Facility (www.dirac.ac.uk). The DiRAC component of CSD3 at Cambridge was funded by BEIS, UKRI and  STFC capital funding and STFC operations grants. DiRAC is part of the UKRI Digital Research Infrastructure. 
We are grateful to the CSD3 support staff for assistance.
Funding for this work came from STFC grant ST/T000945/1. SR is supported by UKRI Stephen Hawking Fellowship EP/W005433/1.

\appendix
\section{Range of accessible $m_a$ values on the lattice}\label{appendix:reach}
\begin{figure}
  \includegraphics[width=0.95\linewidth]{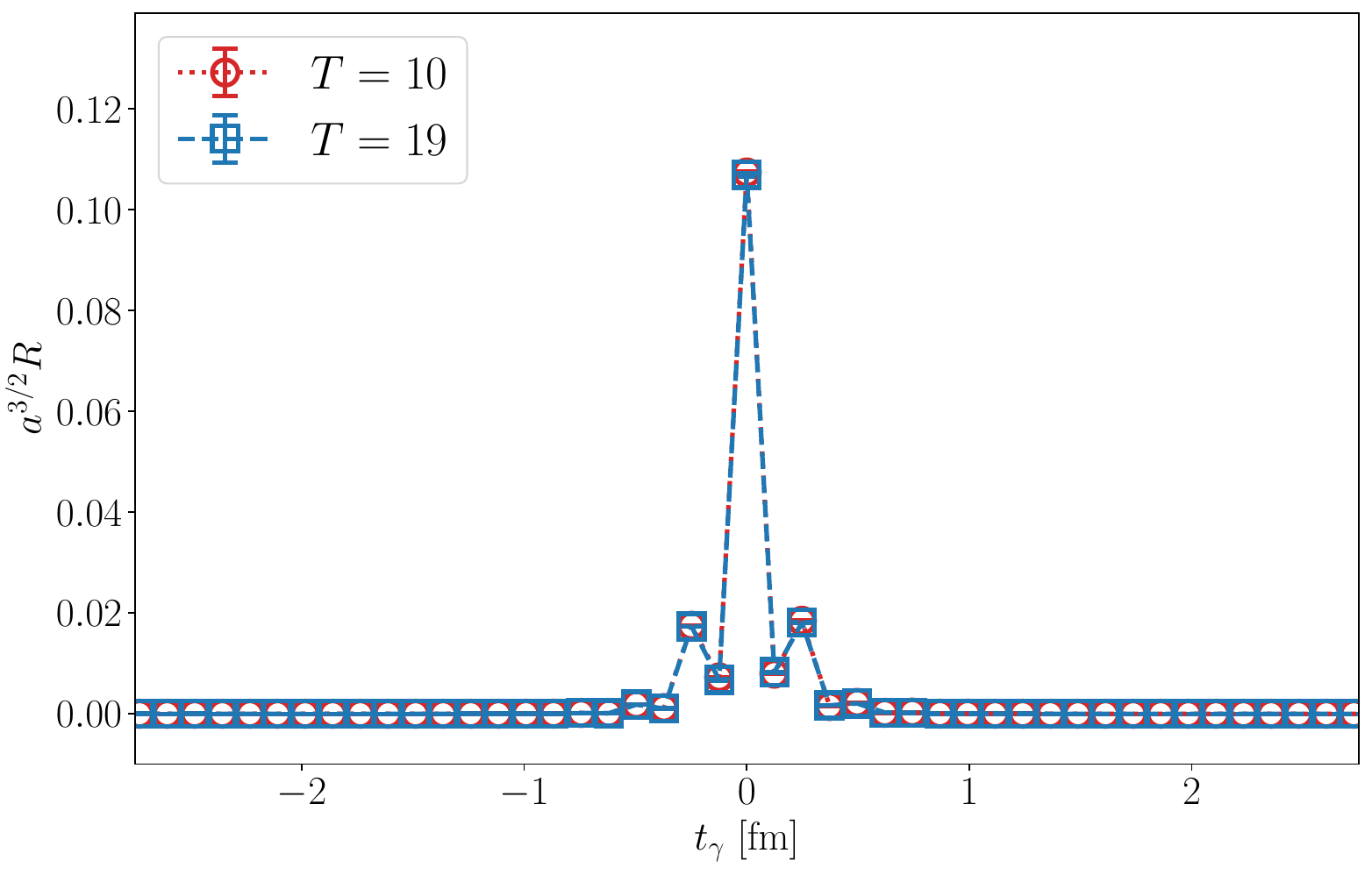}
  \caption{Normalised summand in lattice units for the case of an $m_a=0\GeV$ ALP plotted as a function of $t_{\gamma}$, the time separation between $j_a$ and $j_{\gamma}$ (see Figure~\ref{fig:3pt-pic}) in fm. Results are shown for two values of T (the time position  of $\mathcal{O}_{J/\psi}$) corresponding to 10 ($=1.24\,\mathrm{fm}$) and 19 ($=2.36\,\mathrm{fm}$) on the c-5 ensemble. The oscillations are a consequence of using a staggered quark formalism. }
  \label{fig:summand_0GeVaxion}
\end{figure}

One way to monitor the lattice QCD calculation is to plot the normalised summand that enters the determination of $\tilde{C}_{2pt}(T)$ from $C_{3pt}(t,T)$ using Eq.~\eqref{eq:3to2}. 
The normalised summand is defined as 
\begin{equation}
\label{eq:summand}
R(t_\gamma)= \frac{e^{-\omega t_{\gamma}} C_{3pt}(t_{\gamma},T)}{a_0 e^{-M_0T}} \, ,
\end{equation} 
for each value of $T$ used. Here the ground-state parameters $a_0$ and $M_0$ come from the correlator fit and they remove the exponential dependence of the summand on $T$ so that different $T$ values can be compared directly to each other. 

At low ALP masses, the summand is very strongly peaked around $t_{\gamma} \approx 0$, i.e. where the vector (coupling to photon) and pseudoscalar (coupling to ALP) current insertions coincide in time. This is illustrated in Fig.~\ref{fig:summand_0GeVaxion} for the case of an ALP of zero mass on the c-5 ensemble. It is then easy to arrange for the position of the vector current coupling to the $J/\psi$ to be outside this time region. Indeed in the case illustrated in Fig.~\ref{fig:summand_0GeVaxion} we use $T$ values ranging from 10 to 19 (see Table~\ref{tab:t_values}) corresponding to 1.24 to 2.36 fm. We see from Fig.~\ref{fig:summand_0GeVaxion} that all of the $T$ values lie well outside the peak of the summand and the normalised summand is independent of the $T$ value. This means that there is plenty of lattice time for the 3-point function to project out the charmonium vector states before the operators coupling to $\gamma$ and $a$ are reached. The resulting 2-point correlation function $\tilde{C}_{2pt}$ then has a `clean' multi-exponential form with amplitudes that are the product of the amplitude to form a particular vector charmonium state from the vacuum and for it to decay to $\gamma a$. We are only interested in the vector ground-state here, but we fit to a multi-exponential form to remove excited-state contamination from the ground-state parameters. 

The summand is strongly peaked in the case of Fig.~\ref{fig:summand_0GeVaxion}, and independent of $T$, because once the $J/\psi$ emits a photon with large energy (close to $M_{J/\psi}/2$) the resulting charmonium system is pushed far off-shell and decays exponentially fast in $t_{\gamma}$. 

\begin{figure}
  \centerline{
    \includegraphics[width=0.95\linewidth]{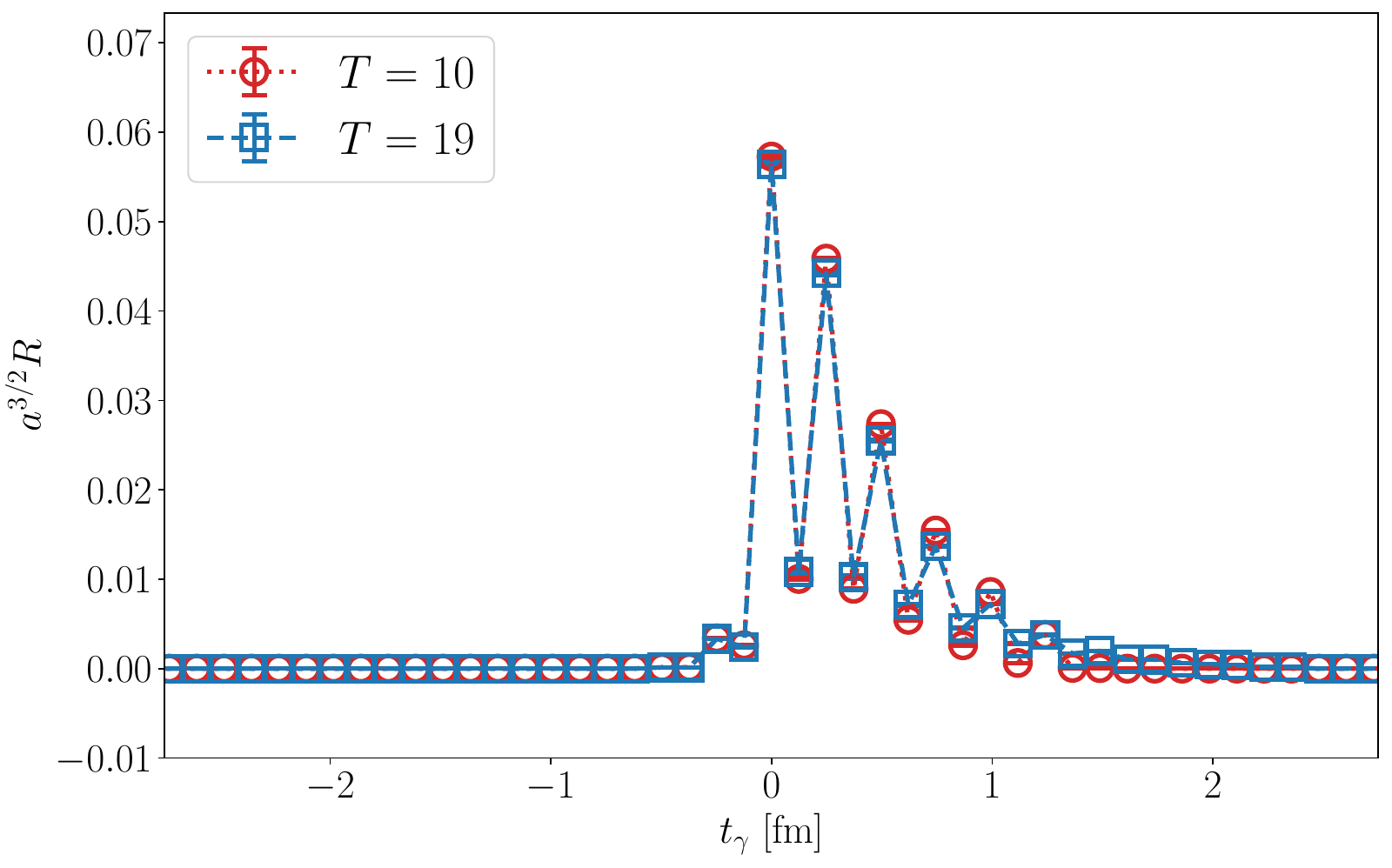}
  }
  \caption{As for Fig.~\ref{fig:summand_0GeVaxion} but for the case of an ALP with  $m_a=2.5\GeV$.}
  \label{fig:summand_2.5GeVaxion}
\end{figure}

\begin{figure}
  \centerline{
    \includegraphics[width=0.95\linewidth]{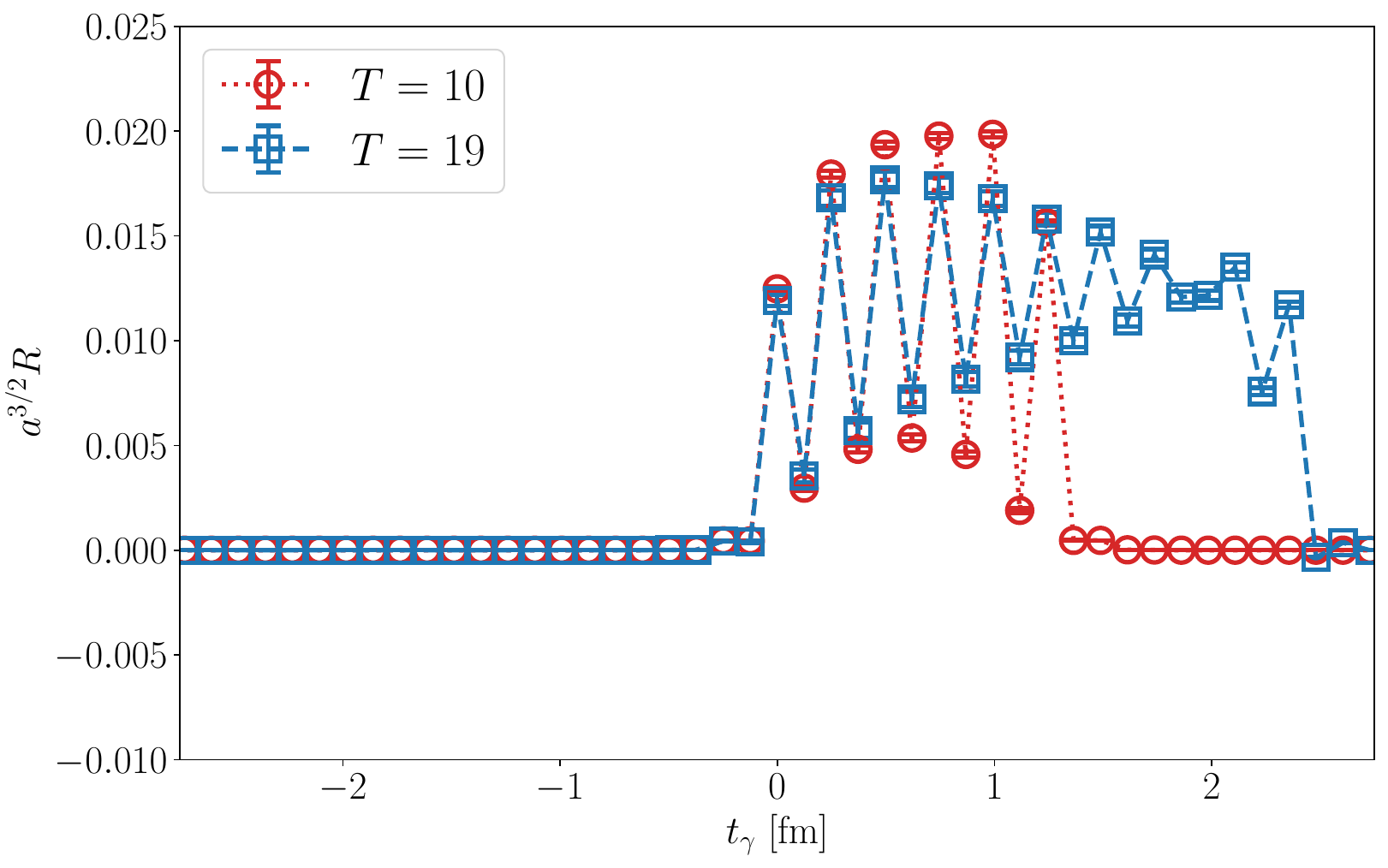}
  }
  \caption{As for Fig.~\ref{fig:summand_0GeVaxion} but for the case of an ALP with  $m_a=2.985\GeV$. Notice that the normalised summand now extends up to the value of $T$. }
  \label{fig:summand_2.985GeVaxion}
\end{figure}

The picture changes as the ALP mass grows to be closer to that of the $J/\psi$. Consider the situation where the $j_{\gamma}$ (photon) operator is between the $\mathcal{O}_{J/\psi}$  and $j_a$ (ALP) operators in time (see Fig.~\ref{fig:3pt-pic}) . Now the emitted photon has smaller energy and the intermediate state between the photon and ALP operators is closer to being on-shell and can `live' for a longer time. This means that the summand is no longer as sharply peaked but becomes asymmetric, with a longer time extent out in the direction towards the position of the $J/\psi$ operator. When the photon operator is on the other side of the ALP operator the intermediate state is again off-shell so the summand has a shorter time extent in that direction. Fig.~\ref{fig:summand_2.5GeVaxion} illustrates this case for an ALP mass of 2.5 GeV which is $0.81M_{J/\psi}$. Now the peak extends on the positive side of the origin to a point close to our smallest $T$ value of 1.24 fm. The normalised summand is still independent of $T$. This case still allows for accurate results for the ground-state amplitude and the form factor because we use multiple $T$ values, extending to 2.36 fm for this case. 

Pushing the ALP mass too close to the $J/\psi$ mass, however, results in 2-point correlation functions which are difficult to fit and which do not allow a form factor value to be extracted. This is illustrated in Fig.~\ref{fig:summand_2.985GeVaxion} for an ALP mass of 2.985 GeV, close to the $\eta_c$ mass and $0.96 M_{J/\psi}$. Now the intermediate state between the photon and ALP operators is very close to being on-shell and can live for a long time, limited only by the position of $\mathcal{O}_{J/\psi}$. The normalised summand now depends on $T$ because the `peak' structure extends up to  $T$. We do not use such large values of $m_a$ in our calculation. These would in any case be close enough to $M_{\eta_c}$ for the ALP and $\eta_c$ to mix, an effect which is not included in our calculation (nor is it included in the perturbative calculations).

\bibliography{axion}{}
\bibliographystyle{apsrev4-2}

\end{document}